\documentstyle[twoside]{article}

\input{psfig}
\def\Journal#1#2#3#4{{#1} {\bf #2}, #3 (#4)}

\def\NPB{{\em Nucl. Phys.} B}
\def\PLB{{\em Phys. Lett.}  B}
\def\MPLA{{\em Mod. Phys. Lett.}  A}
\def\IJMP{{\em Int. J. Mod. Phys.}  A}

\def\JHEP{\em J. High Energy Phys.}
\def\PRD{{\em Phys. Rev.} D}

\def\ep{\epsilon}

\def\be{\begin{equation}}
\def\ee{\end{equation}}
\def\bea{\begin{eqnarray}}
\def\eea{\end{eqnarray}}

\begin{document}


\title{MAGNETIC $Z_N$ SYMMETRY IN 2+1 DIMENSIONS.}

\author{ A. Kovner \\Department of Mathematics and Statistics,
University of Plymouth,\\ Plymouth PL8 4AA, UK}

\maketitle

\abstract{This review describes the role of magnetic symmetry in
2+1 dimensional gauge theories. In confining theories without
matter fields in fundamental representation the magnetic symmetry
is spontaneously broken. Under some mild assumptions, the
low-energy dynamics is determined universally by this spontaneous
breaking phenomenon. The degrees of freedom in the effective
theory are magnetic vortices. Their role in confining dynamics is
similar to that played by pions and $\sigma$ in the chiral
symmetry breaking dynamics.
 I give an explicit derivation
of the effective theory in (2+1)-dimensional weakly coupled
confining models and   argue that it remains qualitatively the
same in strongly coupled (2+1)-dimensional gluodynamics.
Confinement in this effective theory is a very simple classical
statement about the long range interaction between topological
solitons, which follows (as a result of a simple direct classical
calculation) from the structure of the effective Lagrangian. I
show that if fundamentally charged dynamical fields are present
the magnetic symmetry becomes local rather than global. The
modifications to the effective low energy description in the case
of heavy dynamical fundamental matter are discussed. This
effective lagrangian naturally yields a bag like description of
baryonic excitations. I also discuss the fate of the magnetic
symmetry in gauge theories with the Chern-Simons term. }

  \vspace{1cm}

\tableofcontents
\newpage

\section{Introduction}

This review  is devoted to detailed discussion of the magnetic $Z_N$ symmetry
in 2+1 dimensional $SU(N)$ gauge theories.
The main motivation behind it, is
understanding
confinement in gluodynamics in terms of universal properties determined
by realization of global symmetries.

Understanding of the low energy structure of strongly interacting
theories is a very difficult matter. There are instances however
when we can understand main features of the low-energy dynamics
without being able to solve the ``microscopic'' theory in detail.
This happens whenever we are lucky enough to have a spontaneously
broken global continuous symmetry. The Goldstone theorem assures
us that such a spontaneous symmetry breaking results in the
occurrence of massless particles. The latter are, of course,
natural low-energy degrees of freedom. Moreover, the original
symmetry must be manifest in the effective theory that governs the
interactions between the Goldstone bosons, albeit its
implementation is nonlinear. This severely constrains the
interactions of the Goldstone bosons and, in fact, lends
considerable predictive power to the effective theory.

The classic example of this type is the spontaneous
chiral symmetry breaking   which
results in the
 the low-energy physics being dominated
by pions which are described by an effective chiral  Lagrangian.
In real life the chiral symmetry is broken
explicitly,  the pions are massive. However, the effects due to the small
explicit breaking are easily accommodated in the chiral Lagrangian.
The great thing about the chiral Lagrangian, of course, is that it is
universal.
It does not care what exact dynamics is responsible for the symmetry
breaking, what are the degrees of freedom of the ``microscopic'' theory
or any other fine details. All you need to know is that there was a
symmetry, and this symmetry was spontaneously broken.

Can some universal considerations of a similar kind
determine the structure of low-energy theory in pure
gluodynamics?
The chiral symmetry  is of no relevance
here. Nevertheless, I
will argue that the symmetry path is a very fruitful one. The main
thesis is this: there is a global discrete symmetry in gluodynamics,
which is spontaneously broken in the vacuum.
In what follows this symmetry will be called the magnetic $Z_N$ symmetry.
Although it is
discreet and, therefore, does not have all the bliss of the
Goldstone theorem, under some natural assumptions it does indeed
determine the low-energy dynamics. In physical (3+1)-dimensional theory
this symmetry is of a somewhat unusual type -- its charge is not a volume
integral, but rather a surface integral. The implications of such a
symmetry for the low-energy Lagrangian have to be studied in more
detail; this has not been done so far. However, one can go much further
in 2+1 dimensions, where this symmetry is of a familiar garden
variety. This  review will,
therefore,  be devoted to
(2+1)-dimensional non-Abelian theories.
The structure of the paper is the following. In Sec. 2
I will discuss in detail the
nature of the magnetic $Z_N$ symmetry and the explicit construction of
both, the generator of the group and the local order parameter.
In Sect. 3 I will show how this symmetry is realized in different
phases of  Abelian gauge theories. I will show that in the Coulomb
phase this symmetry is spontaneously broken and in the Higgs one
 it is
respected by the vacuum state. In the Abelian theories the magnetic
symmetry is presented by a continuous  U(1)  group -- its spontaneous
breaking leads to the appearance of a massless particle, the photon.
I will derive the low-energy effective
Lagrangian that describes this symmetry breaking pattern and show that
it exhibits logarithmic confinement in the Coulomb phase.
In Sec. 4 I will discuss non Abelian confining
theories. Here the magnetic symmetry is
discrete, but it is also spontaneously broken as in the Coulomb
phase of QED. In weakly interacting theories, like the Georgi-Glashow model,
 it is possible to derive
an effective Lagrangian which follows from this symmetry breaking
pattern. I will demonstrate that, due to the fact that the
magnetic symmetry is discrete, the effective theory exhibits
linear confinement, and that this confinement mechanism in the
effective theory appears very simply at the classical level. I
will give arguments that the basic structure of the effective
Lagrangian as well as   physics of confinement stay  the same in
the limit of pure gluodynamics. I will discuss similarities and
differences between the confining properties of the weakly
interacting non-Abelian theories and gluodynamics. In Sec. 5 I
will discuss the modifications to the magnetic symmetry concept
when fundamentally charged matter fields are present. It turns out
that in theories of this kind the magnetic $Z_N$ symmetry is
gauged. Consequently the low energy effective theories is a $\it
local$ $Z_N$ invariant model. It is however still quite useful. In
particular this approach yields a distinctive low energy
description of baryonic states as "bags" of deconfined phase which
contain the fundamental quarks inside them. Sec. 6 is devoted to
the discussion of $SU(N)$ gauge theories with the Chern-Simons
term. I show that the magnetic symmetry is unbroken in these
theories in the perturbative domain. It is however possible that
it is spontaneously broken at low values of the Chern-Simons
coefficient, and the theories are then confining. I end with a
short discussion in Sec.7.

\section{The magnetic symmetry in 2+1 dimensions}

A while ago 't Hooft gave an argument establishing
that a non-Abelian SU$(N)$ gauge theory
without charged fields in the fundamental representation
possesses a global $Z_N$ symmetry.\cite{thooft}

Consider first a theory with several adjoint Higgs fields, so that
by varying
parameters in the Higgs sector
the SU$(N)$ gauge symmetry can be broken completely. In
this phase the perturbative spectrum will contain the usual massive
``gluons'' and the  Higgs particles. However,
 in addition to those, there will
be heavy stable magnetic vortices. Those are analogs of
Abrikosov-Nielsen-Olesen vortices in the superconductors; they
must be stable by virtue of the following topological argument.
All the fields in the vortex configuration away from the vortex core are
  pure gauge,
\begin{equation}
H^{\alpha}(x)=U(x)h^{\alpha}, \qquad  A^\mu=iU\partial^\mu\,
U^\dagger .
\end{equation}
Here the index $\alpha$ labels the scalar fields in the theory,
$h^\alpha$ are the constant vacuum expectation values of these fields, and
$U(x)$ is a unitary matrix. As one goes around the location of
the vortex in space, the matrix $U$ winds nontrivially in the gauge
group. This is possible, since the gauge group in the theory without
fundamental fields is SU$(N)/Z_N$. It has a nonvanishing first
homotopy group $\pi_1({\rm SU}(N)/Z_N)=Z_N$. Practically,
this means that when one makes a
  full circle around the vortex location, $U$  does not
return to
the same SU$(N)$ group element $U_0$, but, rather, ends up at
$\exp\{i{2\pi\over N}\}U_0$. Adjoint fields do not feel this type of
discontinuity in $U$ and, therefore, the energy of such a configuration
is finite. Since such a configuration can not be smoothly deformed
into a trivial one, a single vortex is stable. Processes involving
annihilation of $N$ such
vortices in the vacuum are allowed since the $N$-vortex configurations
are topologically trivial.
One can, of course, find explicit vortex solutions once the Higgs
potential is specified.
As any other semiclassical solution at weak coupling,
its  energy   is inversely proportional to the
gauge coupling constant and is, therefore, very large.
One is  faced with the situation where the spectrum of the
theory contains a stable particle even though its mass is much larger than
the masses of many other particles (the gauge and Higgs bosons)
and the phase space for its decay into
these particles is enormous. The only possible reason for the existence of
such a heavy stable particle is that it  carries a conserved quantum
number.
The theory must, therefore,  possess a global symmetry which is unbroken
in the completely Higgsed phase. The symmetry group must be $Z_N$
since the number of vortices is only conserved modulo $N$.
't Hooft dubbed this symmetry ``topological.''   I prefer to call
it ``magnetic'' for reasons that will become apparent in a short while.

Now imagine smoothly
changing  the parameters in the Higgs sector so
that the expectation values of the Higgs fields become
 smaller, and, finally, the theory undergoes a phase transition
into the
confining phase. One can further change the parameters so that the
adjoint scalars become heavy and eventually decouple completely from
the  glue. This limiting process does not change   topology of the gauge
group and, therefore, does not change the symmetry content of the
theory. One concludes that the pure Yang-Mills theory also
possesses a $Z_N$ symmetry.
Of course since the confining phase is separated from the completely
Higgsed phase by a phase transition one may expect that the $Z_N$
symmetry in the confining phase is realized differently in the vacuum
than in the completely ``Higgsed" phase. In
fact,
the original paper by 't Hooft, as well as subsequent work,\cite{zn}
convincingly
argued that in the confining phase the $Z_N$ symmetry is spontaneously
broken, the  breaking being related to the confinement phenomenon.
We will have more to say about this later.

The physical considerations given above can be put on a firmer formal
basis. In particular, one can explicitly
construct\,\cite{kovner1}  the $Z_N$ generator
as well as the order parameter associated with it,
the operator that creates the
magnetic vortex.
We start this discussion by considering Abelian theories, where things
are simpler and are   under full control.

\subsection{Abelian theories}

In the  U(1)   case the homotopy group is $Z$ and, therefore,
the magnetic symmetry is  U(1)  rather than $Z_N$.
Identifying the relevant charge
is, in fact, an absolutely straightforward task. It is nothing but
the magnetic flux through the equal time plane, with the associated
conserved current being the dual of the electromagnetic field strength
\begin{equation}
\Phi=\int d^2x B(x)\, ,\qquad \partial^\mu\tilde F_\mu=0\,.
\label{phiqed}
\end{equation}

The current conservation is insured by the Bianchi identity.
It may come as a surprise that we are seriously
 considering  a current whose
conservation equation is an ``identity.'' However ``identity'' is a
relative thing. The conservation equation is trivial only because we have
written the
components of the field strength tensor in terms of the vector
potential. The introduction of the vector
potential is none other than the
 potentiation of the conserved current $\tilde
F_\mu$,
that is explicit solution of the conservation equation. In exactly
the same way
one can potentiate any conserved current, and such a potentiation will
turn the pertinent conservation equation into identity. Thus $\tilde
F_\mu$
has exactly the same status as any other
local conserved gauge invariant current, and should be treated as
such.

Once we have the current and the charge, we also know the elements of the
symmetry group.
A group element of the  U(1)  magnetic symmetry group is
\begin{equation}
W_\alpha(\infty)=\exp\{i\alpha\Phi\}
\end{equation}
for the arbitrary value of $\alpha$.
The notation $W$ is not accidental here, since the group element is
indeed a large spatial Wilson loop
defined on a contour that encloses the whole system.

The question that might bother us is whether this group
acts   nontrivially on any of the  local physical observables in our theory.
The obvious gauge invariant observables like $B$ and $E$ commute
with $W$. However,
there is another set of local gauge invariant observables in
the theory which do indeed transform nontrivially under the action of $W$.
Consider, following 't~Hooft, the operator  of the ``singular gauge
transformation,"
\begin{equation}
V(x)=\exp {\frac{i}{g}}\int d^2y\
\left[\epsilon_{ij}{\frac{(x-y)_j}{(x-y)^2}}
E_i(y)+\Theta(x-y)J_0(y)\right]  \,,
\label{vqed1}
\end{equation}
where $\Theta(x-y)$ is the polar angle function and $J_0$ is the
electric  charge density of whatever matter fields
are present in the theory.
The cut discontinuity in the function $\Theta$
looks bothersome; but it  is, in fact, not physical and
completely harmless. The gauge function jumps across the discontinuity
by $2\pi/g$, but since the only dynamical fields in the theory have
charges that are integer multiples of $g$,
the discontinuity is not observable.  The cut
 can be chosen parallel to the horizontal axis.
Using  the Gauss'  law constraint this can be cast in a different form,
which we   find more convenient for our discussion,
\begin{equation}
V(x)=\exp {\frac{2\pi i}{g}}\int_C dy^i\
\epsilon_{ij}E_i(y)\,,
\label{vqed}
\end{equation}
where the integration goes along the cut of the function $\Theta$
which starts at the point $x$ and goes to spatial infinity.
In this form it is clear that the
operator does not depend on where precisely one chooses the
cut to lie.  To see this, note that
changing the position of the cut $C$ to $C'$ adds
$${2\pi\over g}\int_{S}d^2x \partial_iE^i$$
 to the phase  (here $S$ is the area
bounded  by $C-C'$). In the theory under consideration only
charged particles where charges multiples of $g$ are present.
Therefore, the charge within any closed area is a multiple integer
of the gauge coupling $ \int_{S}d^2x \partial_iE^i=gn$ and the
extra phase factor is always unity. The only point in space where
the action of $V(x)$ on any physical state is nontrivial is the
point $x$. Therefore,  the field $V(x)$ acts like any other local
field. With a little more work one can prove~\cite{kovner} not
only that $V(x)$ is a local field, in the sense that it commutes
with any other local gauge invariant operators $O(y)$,  $x\ne y$,
but also that it is a {\em bona fide} Lorentz scalar.

The physical meaning of the operator $V$ is very simple. Calculating
its commutator with the magnetic field $B$ we find
\begin{equation}
V(x)B(y)V^\dagger(x)=B(y)+{2\pi\over g}\delta^2(x-y)\,.
\end{equation}
Thus,
$V$ creates a pointlike magnetic vortex of flux
$2\pi/g$.
This commutator also tells us that
\begin{equation}
W^\dagger_{\alpha}V(x)W_\alpha=e^{i{2\pi\alpha\over g}}V(x)\,,
\end{equation}
and, therefore,  $V$ is indeed a local eigenoperator of the magnetic  U(1)
symmetry group.

Equations (\ref{phiqed}) and (\ref{vqed}) formalize the physical arguments
of 't Hooft
in the Abelian case.
We have explicitly constructed the generator and the local order
parameter of the magnetic symmetry.

\subsection{Non-Abelian theories at weak coupling}

Let us now move on and consider an analogous construction for
the non-Abelian
theories. Ultimately we are
interested in   pure Yang-Mills theory.  However,  it is  quite
illuminating to start with the theory with an adjoint Higgs field and
take the decoupling limit   later. For simplicity we
discuss the  SU(2) gauge theory.
Consider the
Georgi-Glashow model -- SU(2)  gauge theory with the adjoint Higgs
field,
\begin{equation}
{\cal L}=-\frac{1}{4}F_{\mu \nu }^{a}F^{a\mu \nu }+\frac{1}{2}({\cal D}_{\mu
}^{ab}H^{b})^{2}+\tilde{\mu}^{2}H^{2}-\tilde{\lambda}(H^{2})^{2}\,,
\label{lgg}
\end{equation}
where
\begin{equation}
{\cal D}_{\mu }^{ab}H^{b}=\partial _{\mu }H^{a}-gf^{abc}A_{\mu }^{b}H^{c}\,.
\end{equation}

At large and positive $\tilde{\mu}^{2}$ the model is weakly coupled.
The  SU(2)  gauge symmetry is broken down
to  U(1)  and the Higgs mechanism takes place. Two gauge bosons, $W^{\pm }$,
acquire a mass, while the third one, the ``photon,"  remains massless
to all orders in perturbation theory.
The theory in this region of the parameter space very much resembles
 electrodynamics
of the  vector charged field. The Abelian construction can, therefore,
be repeated.
The  SU(2)  gauge invariant analog of the conserved dual field strength is
\begin{equation}
\tilde{F}^{\mu }={1\over 2}[\epsilon_{\mu\nu\lambda}F^a_{\nu\lambda}n^a-
\frac{1}{g}\epsilon ^{\mu \nu \lambda}
\epsilon ^{abc}n_{a}({\cal D}_{\nu }n)^{b}({\cal D}_{\lambda }n)^{c} ]\,,
 \label{F}
\end{equation}
where $n^a\equiv{\frac{H^a}{|H|}}$ is the unit vector in the
direction of the Higgs field.
Classically
this current satisfies the conservation equation
\begin{equation}
\partial^\mu\tilde F_\mu=0\,.
\label{fa}
\end{equation}
The easiest way to see this is to choose a unitary gauge of the form
$H^a(x)=H(x)\delta^{a3}$. In this gauge $\tilde F$ is equal to the
Abelian part of the dual field strength in the third direction in the
color space. Its conservation then follows from the Bianchi
identity.

Thus, classically the theory has a conserved  U(1)  magnetic charge
$\Phi=\int d^2x \tilde F_0$,  just like QED.
However the unitary gauge can not be imposed at the points where $H$
vanishes,
which necessarily happens in the core of the  't Hooft-Polyakov
monopole. It is well-known, of course,\cite{polyakov} that the
monopoles are the
most important nonperturbative configurations in this model. Their presence
leads to a nonvanishing small photon mass,
as well as
  confinement of the charged gauge bosons, with a tiny
nonperturbative string tension. As far as the monopole effects on
the magnetic flux are concerned, their presence leads to a quantum
anomaly in the conservation equation (\ref{fa}). As a result,
 only
the discrete $Z_2$ subgroup
of the transformation group generated by $\Phi$
remains
unbroken in  quantum theory. The detailed discussion of this
anomaly, the residual $Z_2$ symmetry  and their
relation to monopoles is given in Ref. 15.
The nonanomalous $Z_2$ magnetic
symmetry transformation is generated by the operator
\begin{equation}
U=\exp\{i{g\over 2}\Phi\}\,.
\label{u}
\end{equation}
The order parameter for the magnetic $Z_2$ symmetry is constructed
analogously to QED as a singular gauge transformation generated by the
 gauge invariant electric charge operator,
\begin{equation}
J^\mu=\epsilon^{\mu\nu\lambda}\partial_\nu (\tilde F^a_{\lambda}n^a),
\ \ \ \ Q=\int d^2x J_0(x)  \,.
\label{qqcd}
\end{equation}
Explicitly
\begin{eqnarray}
V(x)&=&\exp {\frac{i}{g}}\int d^2y\ \left[\epsilon_{ij}{\frac{(x-y)_j}{(x-y)^2}
} n^a(y)E^a_i(y)+\Theta(x-y)J_0(y)\right]  \nonumber\\[0.2cm]
&=&\exp {\frac{2\pi i}{g}}\int_C dy^i\,
\epsilon_{ij}n^aE^a_j(y).
\label{VQCD}
\end{eqnarray}
One can think of it as of
a singular  SU(2)  gauge transformation with the field
dependent gauge function
\begin{equation}
\lambda^a(y)={\frac{1}{g}}\Theta(x-y)n^a(\vec y)  \,.
\label{lambda}
\end{equation}
This field dependence of the gauge function ensures the gauge
invariance of the operator $V$.
Just like in QED, it can be shown~\cite{kovner1,kovner3} that
the operator $V$ is a local scalar field.
Again like in QED, the vortex operator $V$ is a local eigenoperator of
the Abelian magnetic field $B(x)=\tilde{F}_{0}$.
\begin{equation}
\lbrack V(x),B(y)\rbrack=-{\frac{2\pi }{g}}V(x)\delta ^{2}(x-y) \,.
 \label{com}
\end{equation}
That is to say, when acting on a state it creates a pointlike magnetic vortex
which carries a quantized unit of the magnetic flux.

The magnetic $Z_2$ acts
on the vortex field $V$ as a phase rotation by $\pi$
\begin{equation}
e^{i{g\over 2} \Phi }V(x)e^{-i{g\over 2} \Phi }=-V(x)\,.
\label{magntr}
\end{equation}
This is an explicit realization of the magnetic $Z_2$ symmetry in
the Georgi-Glashow model.

\subsection{Gluodynamics}

From the Georgi-Glashow model we can easily get  to   pure
Yang-Mills theory.
This is achieved by smoothly varying the
$\tilde\mu^2$ coefficient in the Lagrangian, so that it
becomes negative and eventually arbitrarily large.
In this limit the Higgs field has a large mass and, therefore, decouples
leaving pure gluodynamics behind.
It is well known\,\cite{fradkin} that
in this model
the weakly coupled Higgs regime and
the strongly coupled
confining regime are not separated by a phase
transition. The  limit of pure Yang-Mills in this model is, therefore, smooth.

In the  limit of pure Yang-Mills the expressions
(\ref{F}), (\ref{VQCD}), and (\ref{u})
have to be taken with care.\cite{chris}
When the mass of the Higgs field is very large, the
configurations  that dominate the path integral of the theory
are those with a very small value of the
modulus of the Higgs field $|H|\propto 1/M$.
The modulus of the Higgs field, in turn, controls   fluctuations of
the unit vector $n^a$, since the kinetic term for $n$ in the Lagrangian
is $|H|^2(D_\mu
n)^2$. Thus, as the mass of the Higgs field increases,
 the fluctuations of $n$
grow both, in  amplitude and frequency,
 and the magnetic field operator $B$ as defined
in Eq.~(\ref{F}) fluctuates wildly. Of course,
this situation is  not unusual. It happens whenever one
wants to consider (in the effective low-energy theory) an operator which
explicitly depends on fast, high-energy variables. The standard way of
 dealing
with it is to integrate over the fast variables.
There could be two possible outcomes of this ``integrating out"
procedure. Either the
operator in question becomes trivial (if it depends strongly on the
fast variables), or its reduced version is well-defined and regular in
the low-energy Hilbert space.
The ``magnetic field'' operator $B$ in Eq.~(\ref{F}) is obviously of
the first type. Since in the limit of pure Yang-Mills  all
 orientations of $n^a$ are equally probable, integrating over the
Higgs field at fixed $A_\mu$ will lead to vanishing   $B$. However,
what interests us is not so much the magnetic field but, rather, the
generator of the magnetic $Z_2$ transformation. It is actually
instructive to consider an operator that performs the $Z_2$
transformation not everywhere in space, but only inside a contour $C$
\be
U_C=\exp \{i{g\over 2}\int_S d^2xB(x)\}\,,
\label{uc}
\ee
with the area $S$ being bounded by $C$.
In the limit of gluodynamics we are led to
consider the operator
\begin{eqnarray}
U_C&=&\lim_{H\rightarrow 0}\int Dn^a\exp\Big\{-|H|^2(\vec D
n_a)^2\Big\}\nonumber\\[0.2cm]
 &\times&\exp\Big\{i{g\over 4}\int_C
d^2x\big(\epsilon_{ij}F^a_{ij}n^a-
\frac{1}{g}\epsilon ^{ij}\epsilon ^{abc}n_{a}({\cal D}_{i }n)^{b}
({\cal D}_{j }n)^{c}\big ) \Big\}\,.
\label{uc1}
\end{eqnarray}
The weight for the $n$ integration  is the
kinetic term for the isovector $n_a$. As was noted
before, the action does not depend on $n^a$ in the Yang-Mills limit since
$H^2\rightarrow 0$. However,
the first term in Eq.~(\ref{uc1})
regulates the path integral and we keep it for this reason.
This operator
may look somewhat unfamiliar, at first sight. However, in a remarkable
paper (see Ref.~18) 
Diakonov and Petrov showed~\footnote{We note that
Diakonov and Petrov had to introduce a regulator
to define the path integral over $n$. The regulator they required
was of precisely
 the same form as in Eq.~(\ref{uc1}).} that Eq.~(\ref{uc1}) is equal to
the trace of the fundamental Wilson loop along the contour
$C$.
\begin{equation}
U_C=W_C\equiv{\rm Tr}{\cal P}\exp \Big\{ig\int_C dl^iA^i\Big\}\,.
\end{equation}

Taking the contour $C$ to run at infinity, we see that in
gluodynamics the generator of the magnetic
$Z_2$ symmetry is the fundamental spatial Wilson loop along the boundary of
the spatial plane.\footnote{There is a slight subtlety here that may be
worth mentioning.
The generator of a unitary transformation should be a unitary
operator.  On the other hand, the trace of the
fundamental Wilson loop  is not unitary. Therefore, one should
strictly speaking consider, instead, a
unitarized Wilson loop $\tilde W={W\over \sqrt{WW^\dagger}}$. However,
the factor
between the two operators $\sqrt {WW^\dagger}$ is an operator that is
only sensitive
to the  field behavior   at infinity. It commutes with all physical
local operators
$O(x)$ unless $x\rightarrow\infty$.
In this aspect,  it is very different from the Wilson loop itself,
which has a nontrivial commutator with vortex operators $V(x)$ at all
values of $x$.
Since the correlators of all gauge invariant local fields in the pure
Yang-Mills theory
are massive and, therefore, short range, the operator
$\sqrt{W(C)W^\dagger(C)}$
at $C\rightarrow\infty$ must be a constant
operator on all finite-energy states. Therefore, the difference between $W$
and $\tilde W$ is
 a trivial constant factor and we will not bother with it in what follows.}

Of course, one does not have to go through the exercise with
the Georgi-Glashow model
in order to show that the fundamental Wilson loop generates a symmetry.
Instead one
can directly consider the commutator
\begin{equation}
[W,H]=\lim_{C\rightarrow\infty}\oint_C dx_i
{\rm Tr}{\cal P} E_i(x)\exp\{i\oint_{C(x,x)} dy_iA^i(y)\}
\rightarrow_{C\rightarrow\infty}0\,.
\label{commutator}
\end{equation}
Here the integral in the exponential on the right-hand side
starts and ends at the point of insertion of the electric field.
For a finite contour $C$ the commutator does not vanish
only along the contour itself --  it does not contain any  bulk
terms. Making the contour $C$ go to infinity and assuming, as usual, that
in the  theory with a finite mass gap at
infinity no physical modes are excited we conclude that the commutator of
$H$ with infinitely large Wilson loop vanishes.\footnote{Note that the
nonvanishing of the commutator at finite $C$ is  of
exactly the same
nature as for any other
``conserved charge,"
which is defined as an integral of local charge density
$Q=\lim_{C\rightarrow\infty}
\int_{|x|\in C}d^2x\rho(x)$. The commutator of such a charge with
the Hamiltonian also
contains surface terms, since the charge density $\rho$ itself never
commutes   with the
Hamiltonian. Rather, the commutator is a total derivative.
For a conserved charge, due to the continuity equation,
this surface term is equal to the circulation of the spatial component
of  the current
$$
[Q,H]=\rightarrow_{C\rightarrow\infty}\oint dx^ij_i\,.
$$
Again, the vanishing of this term  is the consequence of the vanishing
of the physical fields
at infinity in the theory with a mass gap. When the charge is not
conserved,  the commutator contains a bulk term,
in addition to the surface term. It is the
absence  of such  bulk terms that is a unique property of the
conserved charge.
The same conclusion is reached if, rather than considering the
generator
of the algebra, one considers the commutator of the group element for
either continuous or
discrete symmetry groups. Therefore,
the commutator in Eq.~(\ref{commutator}) indeed tells
us that $W$ is a conserved operator.}

Next we consider the vortex operator Eq.~(\ref{VQCD}). Again,
in order to find the limit of pure Yang-Mills
we have to integrate
this expression over
the orientations of the unit vector $n^a$.
This integration is equivalent to averaging over the gauge group.
Following Ref.~18 
one can write  $n_a$ in terms of the SU(2)
gauge transformation matrix $\Omega$,
\begin{equation}
\vec n={1\over 2}{\rm Tr}\Omega\tau\Omega^{\dagger}\tau_3\,.
\label{ucl3}
\end{equation}
The vortex operator in the limit of pure gluodynamics  then becomes
\begin{equation}
\tilde V(x)=\int D\Omega\exp{{2\pi i\over g}\int_C
dy_i\epsilon_{ij}{\rm Tr}\Omega E_j\Omega^{\dagger}\tau_3}\,.
\end{equation}
This form makes it explicit that $\tilde V(x)$ is defined as the
gauge singlet part
of the following, apparently non-gauge-invariant operator:
\begin{equation}
V(x)=\exp {\frac{2\pi i}{g}}\int_C dy^i
\epsilon_{ij}E^3_i(y)
\,.
\label{v33}
\end{equation}
The integration over $\Omega$ obviously projects out the gauge singlet part
of $V$.
In the present case, however, this projection is redundant. This is because
even though $V$ itself is not gauge invariant,
when acting on
a physical state it transforms it into another physical
state.\footnote{This is not a trivial statement, since a generic
non-gauge-invariant operator has
nonvanishing matrix elements between the physical and unphysical sectors.}
By physical states
we mean the states which
satisfy  Gauss' constraint in   pure Yang-Mills theory.
This property of $V$ was noted by 't Hooft.\cite{thooft}
To show this let us consider $V(x)$ as defined
in Eq.~(\ref{v33})
and its gauge transform $V_\Omega=\Omega^\dagger V \Omega$ where
$\Omega$ is
an arbitrary
nonsingular gauge transformation operator.
The wave functional of any physical state depends only on gauge
invariant characteristics
of the vector potential, i.e. only on the values of the Wilson loops over
all  possible  contours,
\begin{equation}
\Psi[A_i]=\Psi[\{W(C)\}]\,.
\end{equation}
Acting on this state by the operators $V$ and $V_\Omega$, respectively,
we obtain
\begin{eqnarray}
V|\Psi\rangle &=&\Psi_V[A_i]=\Psi[\{VW(C)V^\dagger\}] \,,\nonumber \\[0.1cm]
V_\Omega|\Psi\rangle &=&\Psi_V^\Omega[A_i]=\Psi[\{V_\Omega
W(C)V^\dagger_\Omega\}]\,.
\end{eqnarray}
However, it is easy to see that the action of $V(x)$ and $V_\Omega(x)$ on
the Wilson loop is  identical -- they both multiply it by a phase
belonging to the
center of the group
if $x$ is inside $C$, and do
nothing otherwise.
Therefore,
\begin{equation}
V|\Psi\rangle =V_\Omega|\Psi\rangle
\label{prev}
\end{equation}
for any physical state $\Psi$.
Thus, we have
\begin{equation}
\Omega V|\Psi\rangle = \Omega V\Omega^\dagger|\Psi\rangle =
V|\Psi\rangle \,,
\end{equation}
where the first equality follows from the fact that any physical
state is invariant under the
action of any gauge transformation $\Omega$, while the second equality
follows from Eq.~(\ref{prev}).
This equation is nothing but the assertion that the state
$V|\Psi\rangle$ is physical, i.e. invariant under any nonsingular gauge
transformation.

Thus, we have   proved that
when acting on a physical state the vortex operator creates
another physical state.
Given such an  operator, the gauge invariant projection only affects
its matrix elements between unphysical states. Since we are
interested only
in calculating correlators of $V$ between physical states, the
gauge projection is redundant and we can freely use $V$ rather
than $\tilde V$ to represent the vortex operator.

The formulae in this section can be straightforwardly generalized
to the SU$(N)$ gauge theories. Once again one can start with the
Georgi-Glashow-like model, where    SU$(N)$ is Higgsed to ${\rm
U}(1)^{(N-1)}$. The construction of the vortex operator and the
generator of $Z_N$ in this case is very similar; necessary
details are given in Ref.  15. Taking the mass of the Higgs field
to infinity again projects the generator onto the trace of the
fundamental Wilson loop. The vortex operator can be taken as
\begin{equation}
V(x)=\exp\{{ 4\pi i\over gN} \int_C dy^i
\epsilon_{ij}{\rm Tr}(YE_i(y))\}
\label{v2}
\end{equation}
where the hypercharge generator $Y$ is defined as
\begin{equation}
Y={\rm diag} \left(1,1,...,-(N-1)\right)\,,
\end{equation}
and the electric field is taken in the matrix notation $E_i=\lambda^aE^a_i$.
Here
$\lambda^a$ stand for  the SU$(N)$ generator matrices in the fundamental
representation.

To summarize this section,
we have established two important facts. First,
 SU$(N)$ gauge theories in 2+1 dimensions have a global $Z_N$ magnetic
symmetry. The generator of this magnetic symmetry group is the fundamental
Wilson loop around the spatial boundary of the system. Second, this
symmetry has a local order parameter. This
order parameter is a local gauge invariant scalar field which creates
a magnetic vortex of fundamental flux.

The next question we should ask is whether this global symmetry is at all
relevant for low-energy dynamics.
In the next section we will show that this is indeed the case. We will
calculate the expectation
value of $V$ in confining and nonconfining situations and will
show that confinement is rigidly related
to the spontaneous breaking of the magnetic symmetry.

\section{The vacuum realization of the magnetic symmetry, the effective
Lagrangian and confinement. Abelian theory}

Again, we start our discussion with the Abelian theory.
Consider a  U(1)  gauge theory with scalar matter field
\begin{equation}
{\cal L} =  - {1\over 4}F^2
+|D_{\mu} \phi |^2 - M^2 |\phi|^2
- \lambda(\phi^*\phi)^2\,.
\label{lscalar}
\end{equation}
Depending on the values of the coupling constant this theory can be
either  in the
Coulomb phase with the massless photon and logarithmically confined
charges,  or in the
Higgs phase, which is massive -- all  electric charges are screened.

\subsection{Realizations of the magnetic symmetry}

Let us start by calculating $\langle V\rangle $ in the Coulomb phase.
This can be done using the
standard weak coupling perturbation theory.
The expectation value of $V$ is given by the following expression:
\begin{equation}
\langle V(x)\rangle
=N^{-1}\lim_{T\rightarrow\infty}\int dA_0
\langle 0|e^{iTH}
e^{{\frac{2\pi i}{g}}\int_C dy^i
\epsilon_{ij}E_i(y)}e^{i\int A_0[\partial_iE_i-J_0]} e^{iTH}|0\rangle
\,.
\end{equation}
Here $N^{-1}$ is the normalization factor -- the usual vacuum-to-vacuum
amplitude,
$|0\rangle $ is the perturbative Fock vacuum and the integral over $A_0$
is the
standard representation of the projection operator
which projects the Fock vacuum $|0\rangle$ onto the gauge invariant subspace
which satisfies   Gauss' law. As usual, discretizing time,
introducing resolution of identity
at every time slice and integrating over $E_i$ in the phase
space path integral, this expression can be rewritten as a path integral
in the
field space. The result is easy to understand -- it is almost
the same as for the vacuum-to-vacuum amplitude, except that at time
$t=0$ the spatial
derivative of the scalar potential $A_0$ is shifted by a $c$-number field
due to the presence of the vortex operator,

\begin{equation}
\langle V(x)\rangle =N^{-1}\int {\cal D}A_\mu\exp \left[-{1 \over 4}\int d^3y
[\tilde
F_{\mu}(y)-\tilde f_{\mu}(y-x)]^2
+L_{Higgs}\right]\,.
\label{veuc}
\end{equation}
The $c$-number field $\tilde f_{\mu}$ is the magnetic field of an
infinitely thin
magnetic vortex which terminates at the
 point $x$. One can view it as the
Dirac string of a (three-dimensional Euclidean) magnetic monopole,
\begin{equation}
\tilde f_0 = \tilde f_2 = 0\,\qquad
\tilde f_1(y) = {2\pi\over g}\theta(y_1) \delta(y_2) \delta(y_3)
\,.
\end{equation}

Thus, at weak coupling we have to find the solution of the
classical equations of motion following from the action with the
external source Eq.~(\ref{veuc}). The nature of this solution is
clear: it is just the Dirac monopole. The action of this solution
is infrared-finite, since the contribution of the Dirac string
(which normally would be linearly infrared-divergent) is cancelled
by the external source, \be \langle V \rangle =\exp\{-S_{cl}\}\,,
\ee with \be S_{\rm cl}={\Lambda\over g^2}\,. \ee Here $\Lambda$
is an ultraviolet cutoff which has to be introduced since the
action of the pointlike monopole diverges in the ultraviolet. This
ultraviolet divergence is benign since it can be eliminated by the
multiplicative renormalization of the vortex
operator.\cite{polchinski} The important point is that since there
is no divergence in the infrared, the expectation value of $V$ is
nonvanishing. Thus, we conclude that in the Coulomb phase of QED
the magnetic symmetry is spontaneously broken.

The spontaneous breaking of a continuous symmetry must be accompanied by
a massless Goldstone particle. Indeed, in QED such a
particle  exists -- this
is the massless photon. The matrix element of the magnetic current
between the vacuum and the one-photon state is
\be
\langle 0|\tilde F_\mu|k_i
\rangle =Z^{1/2}(0)k_\mu\,,
\ee
where $Z(0)$ is the on-shell photon wave function renormalization.

This is the standard form of the matrix element of a spontaneously broken
current, with $Z(0)$ playing the role of $f^2_\pi$.

Let us now perform the same calculation in the Higgs phase. The
path integral representation in  Eq.~(\ref{veuc}) is still valid. However,
the classical solution that dominates this path integral is now very different.
Since in the Higgs phase the photon has a nonzero
 mass $\mu$ the classical action of the
three-dimensional monopole in the superconducting medium is
linearly divergent in the infrared.\cite{polchinski} Essentially,
the magnetic flux that emanates from the monopole can not spread
out in space (time) but is rather concentrated inside a flux tube
of the thickness $1/\mu$ which starts at the location of the
monopole and goes to a spatial boundary at infinity. The action of
such a field configuration is proportional to the linear size of
the system and diverges in the thermodynamic limit. As a result,
the expectation value of $V$ in the Higgs phase vanishes, \be
\langle V\rangle  =e^{-L}\rightarrow_{L \rightarrow\infty}0\,. \ee
Similarly, the vortex field correlator is given in terms of the
classical energy of a monopole-antimonopole pair in the
superconductor. The Euclidean action of this configuration is
proportional to the distance between the monopole and the
antimonopole and, therefore, the correlator of the $V$ decays
exponentially,
\begin{equation}
\langle
V^* (x)V (y) \rangle \sim e^{-M_V|x-y|}\,,
\end{equation}
with $M_V\propto 1/g^2$ being the mass of the ANO vortex.

This simple calculation can be improved perturbatively.
In the next-to-leading order
one has to calculate the determinant of the Schr\"odinger operator of a
particle in the field of a monopole, and this corrects the value of the
mass $M_V$. Higher orders in perturbation theory can be calculated too;
 we will not pursue this calculation here.

The main lesson is that the expectation value of the order parameter
vanishes in the Higgs phase; thus, the magnetic symmetry is unbroken.

\subsection{The low-energy effective Lagrangian and
the logarithmic confinement of electric charges}

Thus we see that the Coulomb--Higgs phase transition can be described
as  that due to the
restoration of the magnetic  U(1)  symmetry, the pertinent local order
parameter
being the vortex operator $V$.
It should then be true that
the low-energy dynamics in the vicinity of the phase transition
is described by an effective low-energy Lagrangian. For the  U(1)  symmetry
breaking such
an effective Lagrangian can be immediately written as follows:
\begin{equation}
{\cal L}=\partial_\mu V^*\partial^\mu V -\lambda(V^*V-\mu^2)^2
\,.
\label{ldual}
\end{equation}

Although this Lagrangian may seem a bit unfamiliar in the
context of QED, a little thought convinces one
that it
indeed describes all
the relevant light degrees of freedom of the theory.
In the Coulomb phase, where $\langle V \rangle =\mu\ne 0$, the physical
particles are interpolated
by the phase and the radial part of $V$,
\be
V(x)=\rho e^{i\chi}\,.
\ee
The phase $\chi$ is of course the
massless Goldstone boson field, i.e. the photon. The
fluctuation of the radial component $\rho-\mu$  is the lightest scalar
particle,
which in this case is the lightest meson, or scalar positronium.
In the Higgs phase, the field $V$   interpolates physical excitations --
the ANO vortices. Of course, far from the phase transition the vortices
are heavy
and there are other, lighter excitations in the spectrum. The validity of this
effective
Lagrangian on the Higgs phase side is therefore limited to a narrow
critical region where vortices are indeed the lightest particles.

The  charged particles   have not appeared in our discussion so far.
Indeed in 2+1
dimensions the electrical charges are
confined, and,
therefore, we do not expect the charged fields to appear as basic
degrees of freedom in
the effective low-energy Lagrangian. However, our original purpose was
precisely to understand the mechanism of confinement through studying the
effective Lagrangian. Thus, if we are unable to identify the charged objects
in this framework our program is doomed to failure.
Fortunately it is not difficult to understand how the charged states are
represented in the Lagrangian Eq.~(\ref{ldual}). The easiest way to do this
is to identify the electric charge through the Maxwell equation,
\be
J_\mu={1\over 4}\epsilon_{\mu\nu\lambda}\partial_\nu\tilde F_\lambda
\,.
\ee

The dual field strength $\tilde F_\mu$ is
obviously proportional to the conserved U(1)  current
\be
\tilde F_\mu={2\pi\over g}i(V^*\partial V-h.c.)
\,.
\ee
The proportionality constant in this relation is dictated by the fact that
in the Higgs phase the magnetic vortices that carry one unit of the  U(1)
charge, carry the magnetic flux of $2\pi/g$.
The above two relations give
\be
{\frac{g}{\pi }}J_{\mu }=i\epsilon _{\mu \nu \lambda }\partial _{\nu
}(V^{\ast }\partial _{\lambda }V)
\,.
\label{j}
\ee
To calculate the electric charge we integrate the zeroth component of
the current over the two-dimensional plane
\be
{g\over\pi}Q=\mu^2\oint_{C\rightarrow\infty} dx_i\partial_i\chi\,.
\label{q}
\ee
Therefore, the electric charge is  proportional to the winding number of
the phase of the field $V$.

\begin{figure}
\hskip 4cm
\psfig{figure=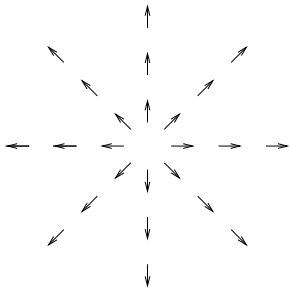,height=1.5in}
\caption{The hedgehog configuration of the field $V$
in the state of unit charge in the Abelian theory.}
\label{fig:hedgehog}
\end{figure}

So, the charged states do appear in the low-energy description in a
very natural way.
A charged state is a soliton of $V$ with a nonzero winding number.
This identification immediately
tells us  that the charged particles are
logarithmically confined. Consider, for example, the
minimal energy configuration
in the sector with the unit winding number.
This is a rotationally invariant
hedgehog, Fig.~\ref{fig:hedgehog},
which, far from the soliton core, has the form
\begin{equation}
V(x)=\mu e^{i\theta (x)}\,.
\end{equation}
Here $\theta (x)$ is the angle between the vector $x$  and one of the
axes.

The self-energy of this configuration is logarithmically divergent in
the infrared due to the contribution of the kinetic term
\begin{equation}
E=\pi \mu ^{2}\ln (\lambda\mu^2L^2)\,.
\label{ech}
\end{equation}
This is nothing but the electromagnetic self-energy
of an electrically charged state associated with the logarithmic
Coulomb potential in two spatial dimensions.
Therefore, the logarithmic confinement of QED is indeed very easily and
transparently seen at the level of the low-energy effective Lagrangian.
In itself this is, perhaps, not such a big  deal, since confinement
in this model
is a kinematic phenomenon: it is a direct consequence of the logarithmic
behavior of the Coulomb potential. We will see later, however, that this
 low-energy picture also naturally  generalizes
to non-Abelian theories and easily accommodates linear confinement.

Before moving on to the non-Abelian theories, let me make two comments. First,
  in perturbative regime the couplings of the effective Lagrangian can be
determined in terms of the couplings of the fundamental QED Lagrangian.
To determine  two constants in Eq.~(\ref{ldual}) one needs two matching
conditions. One of them can be
naturally taken as the coefficient of the infrared logarithm in the
self-energy
of a charged state.
Matching Eq.~(\ref{ech}) with the Coulomb self-energy gives
\be
\mu^2={\frac{g^2}{8\pi^2}}  \,.
\label{couplings1}
\ee
The other coefficient is determined by requiring that the mass of the radial
excitation $\rho$ matches the mass of the scalar positronium, which to
leading order in $g$ is just $2M$.
This condition gives
\be
\lambda={2\pi^2M\over g^2}\,.
\label{couplings2}
\ee
The second comment is about the relation between the vortex operator as
defined in Eq.~(\ref{VQCD})
and the field $V$ that enters the effective Lagrangian
(\ref{ldual}).

The vortex operator, as defined in Eq.~(\ref{VQCD}) has a fixed length,
 whereas
the field $V$ which enters the Lagrangian (\ref{ldual}) is a
conventional complex field. How should one understand that? First of
all, at weak gauge coupling the quartic coupling in the dual Lagrangian
is large, $\lambda \rightarrow \infty $. This condition
freezes the radius of $V$ dynamically. In fact, even at
a finite value of
$\lambda $, if one is interested in the low-energy physics, the radial
component is irrelevant as long as it is much heavier than the phase. Indeed,
at weak gauge
coupling the phase of $V$ (which interpolates the photon) is
much lighter than all   other excitations in the theory. Effectively
 at low energies Eq.~(\ref{ldual}) reduces to a nonlinear $\sigma$
model and one can identify the field $V$ entering Eq.~(\ref{ldual})
directly with the vortex operator of Eq.~(\ref{VQCD}). However, it is
well known that quantum mechanically the radial degree of freedom of the
$\sigma$
model field is always resurrected. The spectrum of such  theory always
contains a scalar particle which can be combined with the phase into a
variable-length complex field. The question is purely quantitative -- how heavy
is this scalar field relative to the phase?

Another way of expressing this is as follows. The fixed length field $V$
is defined at the scale of the ultraviolet cutoff in the original theory.
To arrive
at a low-energy effective Lagrangian one has to integrate over all quantum
fluctuations down to some much lower energy scale. In the process of this
integrating out, the field is ``renormalized'' and acquires a dynamical
radial part. Then the mass of this radial part   is  equal to the mass of
the lowest particle with the same quantum numbers in the original theory.
That is   why the parameters in Eq.~(\ref{ldual}) must be  such that the
mass of the radial part of $V$ would be equal to the mass of the lightest
scalar positronium.

\section{Non-Abelian theories }

In this section I want to extend the construction of the  low-energy
effective Lagrangian
discussed above  to the  non-Abelian theories and see how the realization
of the magnetic symmetry in this Lagrangian is related to confinement. Before
discussing  specific models in detail,
let me present a general argument
which establishes that the spontaneous
breaking of magnetic $Z_N$  in   non-Abelian theories implies the
area law behavior of the fundamental Wilson loop.

\subsection{Broken $Z_N$ means confinement}

As we   discussed in the Sec. 2, the generator of
the magnetic $Z_N$
in pure gluodynamics
is the fundamental Wilson loop around the spatial boundary of the system.
By the same token, the Wilson loop around a closed spatial
contour $C$ generates the
$Z_N$ transformation at the points inside the contour $C$.
Let us imagine that the $Z_N$ symmetry is spontaneously broken in the
vacuum and
consider the expectation value of $W(C)$ in such a state $|0\rangle$ .
The expectation value $\langle 0|W(C)|0 \rangle$ is nothing but the
overlap of the vacuum state $\langle 0|$
and the state $|S\rangle$ which is obtained by acting  on the vacuum by $W(C)$,
 $|S\rangle =W(C)|0\rangle$.
If the symmetry is broken, the wave function $|0\rangle$ depends explicitly
on the degrees of freedom which are non-invariant under the symmetry
transformation, and is peaked around some specific orientation of these
variables in the group space. For simplicity let us think about all
these non-invariant variables as being represented by the vortex field $V$.
In the vacuum state the field
$V$  has a nonvanishing vacuum expectation value (VEV) and
points in some fixed direction in the
internal space. On the other hand, in the state $|S\rangle $  its direction in
the internal space is different -- rotated by $2\pi/N$ -- at the points
inside the area $S$ bounded by $C$, since at these points
the field $V$ has been rotated by
the action of $W(C)$.
In the local theory with finite correlation length the overlap
between the two states approximately factorizes in  the product of the
overlaps taken over the regions of space of linear dimension of the order
of the correlation length $l$,
\begin{equation}
\langle 0|S\rangle =\Pi_x\langle 0_x|S_x\rangle\,,
\label{fact}
\end{equation}
where the label $x$
is the coordinate of the
point in the center of a given small region of space. For $x$
outside the area $S$ the two states $|0_x\rangle $ and $|S_x\rangle $ are
identical and,
therefore, the overlap is unity. However, for $x$ inside $S$ the states
are different and the overlap is a number $e^{-\gamma}$
smaller than unity. The number of such regions inside the area is obviously
of the order of $S/l^2$
and, thus,
\begin{equation}
\langle W(C)\rangle =\exp\{-\gamma{S\over l^2}\}\,.
\end{equation}
Therefore, in the broken phase the spatial Wilson loop
has the area law behavior.

Now consider the unbroken phase. Again the average of $W(C)$
has the form of the overlap
of two states which factorizes as in Eq.~(\ref{fact}). Now, however,
all observables non-invariant under $Z_N$  vanish in the vacuum. The action of the
symmetry generator does not affect the state $|0\rangle $. Hence, the
 state $|S\rangle$  is
  locally exactly the same as the state $|0\rangle$ except along the
boundary $C$.
Therefore, the only regions of space which contribute to the overlap are those
which lay within
one correlation length from the boundary. Thus,
\begin{equation}
\langle W(C)\rangle =\exp\{-\gamma P(C)\}\,,
\end{equation}
where $P(C)$ is the perimeter of the curve $C$.

The crucial requirement for this argument to hold is the existence of a mass
gap in the theory. If the theory contains massless excitations the
factorization
of the overlap does not hold, and so in principle even in the broken phase
the Wilson loop can have a perimeter behavior. This indeed is the case in
the Abelian theories.

We now turn to the discussion of low-energy effective theories.
This will allow us to see an explicit realization of this general argument.

\subsection{The Georgi-Glashow model }

Let us again start with the Georgi-Glashow model.
For simplicity all explicit calculations in this section will be performed for
the SU(2)  gauge theory. Generalization to the  SU(N)  group is not difficult and
is discussed in Ref. 15.

Not much has to be done here to parallel the
calculation of $\langle V\rangle $ of the previous section.
The theory is weakly interacting, and all calculations are explicit. Choosing
the unitary gauge $n^a=\delta^{a3}$, the perturbative calculation becomes
essentially
identical to that in the Coulomb phase of QED. The only difference is that
the charged matter
fields are vectors ($W^\pm_\mu$) rather than scalars ($\phi$), but this only
enters at the level of the loop corrections.
The nonperturbative monopole contributions are there, but they affect
the  value
of $\langle V\rangle$ very little, since $\langle V\rangle\ne 0$ already in
perturbation theory. Thus just like in the Coulomb phase of QED,
$\langle V\rangle\ne
0$ and the magnetic
symmetry
is spontaneously broken.
The real difference comes only when we ask
ourselves what is the effective Lagrangian
that describes the low-energy physics. Here the monopole contributions are
crucial,
since as we have seen before the  U(1)  magnetic symmetry of QED is explicitly
(anomalously)
broken by these contributions to $Z_2$. Therefore the effective Lagrangian
must have  extra terms which reduce the symmetry of the Eq.~(\ref{ldual}).
The relevant effective Lagrangian is
\begin{equation}
{\cal L}=\partial_\mu V^*\partial^\mu V -\lambda(V^*V-\mu^2)^2
+\zeta(V^2+(V^*)^2)\,.
\label{ldualgg}
\end{equation}

The addition of this extra symmetry breaking
term has an immediate effect on the mass of the
``would be" photon -- the phase of $V$. Expanding around $\langle V\rangle
=\mu$ we
see that the phase field now has a mass $m_{ph}^2=4\zeta$. This is
consistent with the classical analysis by Polyakov~\cite{polyakov} --
the monopole
contributions turn the massless photon of QED into a massive (pseudo)scalar
with an exponentially small mass, $m_{ph}\propto\exp\{-4\pi M_W/g^2\}$.
As a matter of fact, for very weak coupling, when the modulus of
$V$ can be considered as frozen, the Lagrangian (\ref{ldualgg}) in
terms of the phase $\chi$ reduces to Polyakov's dual Lagrangian. The exact
correspondence between the two is discussed in Ref. 15.

\begin{figure}
\hskip 3.3cm
\psfig{figure=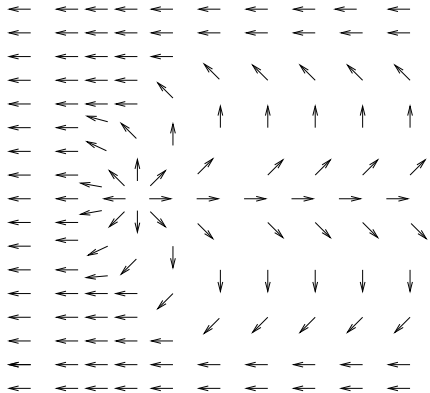,height=2in}
\caption{The  string-like configuration of the field $V$
in the state of unit charge in the presence of
the symmetry breaking terms in the
effective Lagrangian.}
\label{fig:string}
\end{figure}

The explicit symmetry breaking causes a dramatic change in the
topologically charged (soliton) sector.
We know from Polyakov's analysis that the charges in this model
are confined by linear potential,
as opposed to QED where confinement is logarithmic. In the
effective Lagrangian description this is due to the explicit symmetry
breaking term. The crucial point is that the vacuum of the theory is not
infinitely degenerate $\langle V\rangle
=e^{i\chi}\mu$ with arbitrary constant $\chi$,
as in   QED, but only doubly degenerate
$\langle V\rangle
=\pm\mu$. Thus, the lowest energy state in the nontrivial winding
sector cannot be a hedgehog. In the hedgehog configuration the field $V$
at each point at spatial infinity points in a different direction
in the internal space.
This is OK if all these directions are minima of the potential.
Then the total energy of the configuration comes from the kinetic term, and as
we saw, is logarithmic.
However, now the potential has only two minima.
Thus, the hedgehog field is
far from the vacuum everywhere in space. The energy of such a state
diverges as the volume of the system, $E\propto g^{2}m^{2}L^{2}$.
Clearly, to minimize the energy in a state with
a nonzero winding, the system must be   in one of the two vacuum states
in as large a region of space as possible. However, since the field has to wind
when one goes around the position of a soliton even at
an arbitrarily large
distance,
$V$ cannot be aligned with the vacuum everywhere at infinity.
The best bet for a system is,  therefore, to choose
a string-like configuration, Fig.~\ref{fig:string}. The phase of $V(x)$
deviates
from $0$ (or $\pi $) only inside a strip of width $d\sim 1/m_{ph}$ stretching
from the location of the soliton (charge) to infinity. The energy of such a
configuration diverges only linearly with the dimension $L$.
In fact, a back-of-the-envelope estimate with the effective
Lagrangian (\ref{ldualgg}) gives the following  energy
of such a confining string:
$$E\propto g^2m_{ph}L\,.$$
Clearly the energy  of a soliton and an anti-soliton separated
by a large distance $R$ is $E=\sigma R$, with the string tension
$\sigma\propto g^2m_{ph}$.

This is a simple picture of
confinement in the effective Lagrangian approach in the weakly coupled
regime.

The preceding discussion pertains to confinement of ``adjoint"
color charges. So far we have been considering topological solitons
with the unit winding, which corresponds to the charge of the massive $W^\pm$
bosons, or ``massive gluons." It should be noted that
the notion of the adjoint string tension is not an absolute one.
So far, our discussion  neglected the fact that the solitons have a finite
core energy and, therefore, in principle can be created in pairs from the
vacuum.
Thus, the soliton-antisoliton interaction at a distance $R$ can be screened
by creating such a
pair, if $R$ is big enough. The distance at which the string breaking occurs
can be estimated from the energy balance between the energy stored in the
string $E_S=\sigma_{Adj}R$ and the core energy of the soliton-antisoliton pair,
which is twice the mass of the $W$-boson, $2M_W$,  in our model,
\be
g^2m_{ph}R=2M_W\,.
\ee
The distance at which the string breaks is
\be
R_{breaking}\propto{M_W\over g^2}{1\over m_{ph}}\,.
\ee
Since the width of the string is of the order of $1/m_{ph}$, and
$M_W\gg g^2$ at the weak coupling, the length of the string is indeed much
greater
than its width. One can therefore sensibly talk about a
well formed adjoint confining string.

As opposed to the adjoint string tension, the concept of the
fundamental string tension is sharply defined.
This is  because the theory does not contain
particles with fundamental charge and, thus, an external
fundamental charge can not
be screened.
To discuss confinement of external fundamental
charges we have to learn how to deal with half-integer windings.
Imagine adding  some extra
very heavy fields in the fundamental representation
to the Georgi-Glashow model.
The quanta of these fields will carry
a half-integer ``electric" charge
(\ref{q}) and will be confined with a different string tension than $W^\pm$.
To calculate this string tension we should
consider the Abelian Wilson loop with a half-integer charge.
We will now do it in the effective theory framework.

Let us first consider a space-like Wilson loop.
As discussed in the previous section, this operator is closely related to
the generator of the magnetic $Z_2$. In fact, $W(C)$
is nothing but the operator that performs the $Z_2$ transformation inside the
area bounded by the contour $C$.

It is straightforward to write down an operator in the effective theory,
in terms of the field $V$,
that has the same property,
\begin{equation}
W(C)=e^{i\pi \int_{S}d^{2}xP(x)}\,.  \label{wld}
\end{equation}
Here $S$ is the surface bounded by the contour $C$, and $P$ is the momentum
operator   conjugate to the phase of $V$. In terms of the radius and   phase
of $V$ the path integral representation for calculating the vacuum
average of this Wilson loop is
\begin{equation}
\langle W(C)\rangle =\int DV\exp \left\{i\int d^{3}x\rho ^{2}(\partial _{\mu }\chi
-j_{\mu }^{S})^{2}+(\partial _{\mu }\rho )^{2}-U(V)\right\}  \,,
\label{wldpi}
\end{equation}
where $U(V)$ is the $Z_{2}$ invariant potential of Eq.~(\ref{ldualgg}). The
external current $j_{\mu }^{S}(x)$ does not vanish only at points $x$ which
belong to the surface $S$ and is proportional to the unit normal $n_{\mu }$
to the surface $S$. Its magnitude is such that when integrated in the
direction of $n$ it is equal to $\pi $. These properties are
conveniently
encoded in the following expression:
\begin{equation}
\int_{T}dx_{\mu }j_{\mu }^{S}(x)=\pi n(T,C) \,. \label{angle}
\end{equation}
Here $T$ is an arbitrary closed contour, and $n(T,C)$ is the linking
number~\footnote{Note that here we are
dealing
with
the path integral representation, and thus the contour $C$ and the surface $S$
are embedded into a three dimensional Euclidean space. The linking number
between two
curves is also defined in three dimensions.}
between two closed curves $T$ and $C$.

This path integral representation follows immediately if we note that
the conjugate momentum $P$ is $P(x)=2\rho^2(x)\partial_0\chi(x)$ at some
fixed time $t$. This accounts for the linear in $\partial_\mu\chi$ term in the
exponential (\ref{wldpi}). The constant term $j^2$ arises due to
the standard
integration over the conjugate momenta in passing to the path integral
representation.

The path integral representation was constructed for
the  spatial Wilson loops. However, the expression (\ref{wldpi})
is completely covariant, and in this form it is valid for time-like
Wilson loops as well. It is important to note that, although the expression
for the current depends on the surface $S$, the Wilson loop operator
depends only on the contour $C$ that bounds this surface. A simple way to
see this is to observe that a change of variables
$\chi\rightarrow\chi+\pi$
in the
volume bounded by $S+S^{\prime}$ leads to the change $j_\mu^S\rightarrow
j_\mu^{S^{\prime}}$ in Eq.~(\ref{wldpi}). The potential is not affected by
this change since it is globally $Z_2$ invariant. Therefore, the operators
defined with $S$ and $S^{\prime}$ are completely equivalent.

\begin{figure}
\hskip 2.5cm
\psfig{figure=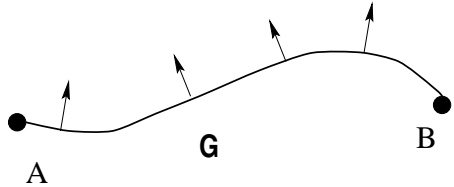,height=1in}
\caption{The external current $j_\mu$
which creates a pair of static fundamental charges
in the effective Lagrangian description.}
\label{fig:current}
\end{figure}

To calculate the energy of a pair of static fundamental charges at
the points $A$ and $B$ we have to consider a time-like fundamental
Wilson loop of
an infinite time dimension. This corresponds to time-independent $j_\mu$ which
does not vanish only along a spatial curve $G$ (in the equal-time
cross section)
connecting the two points and
pointing in the direction normal to this curve, Fig.\ref{fig:current}.
The shape of the curve
itself does not matter, since changing the curve without changing its
endpoints is equivalent to changing the surface $S$ in Eq.~(\ref{wldpi}).

\begin{figure}
\hskip 2cm
\psfig{figure=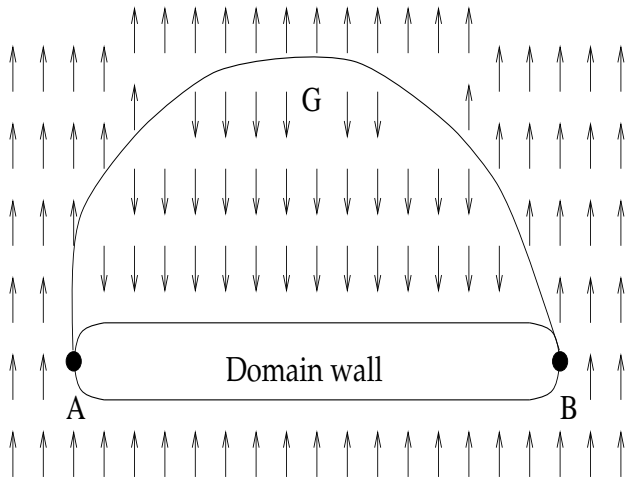,height=2.5in,width=3.2in}
\caption{The minimal energy configuration of $V$ in the presence of a
pair of fundamental charges.}
\label{fig:wall}
\end{figure}

In the classical approximation the path integral (\ref{wldpi}) is
dominated by a static configuration of $V$. To determine it we have to
minimize the energy on the static configurations in the presence of the external
current $j_{\mu }$. The qualitative features of the minimal energy solution
are quite clear. The effect of the external current is to flip the
phase of $V$ by $\pi $ across the curve $G$, as is expressed in
Eq.~(\ref{angle}). Any configuration that does not have this behavior
will  have the
energy proportional to the length of $G$ and to the ultraviolet cutoff scale.
Recall that the vacuum in the theory at hand is doubly degenerate. The
sign change
of $V$ transforms one vacuum configuration into another. Therefore, the
presence of $j_{\mu }$  requires that on the opposite sides of the
curve $G$, immediately adjacent to $G$, there should be different vacuum
states. It is clear, however, that far away from $G$ in either direction the
field should approach the same vacuum state, otherwise the energy of the
configuration diverges linearly in the infrared. The phase of $V$
has to make half a wind somewhere in space to return to the same vacuum
state far below $G$,  as the  vacuum
state that exists far above $G$. If the distance
between $A$ and $B$ is much larger than the mass of the lightest particle in
the theory, this is achieved by having a segment of a domain wall between
the two vacua connecting the points $A$ and $B$. Clearly to minimize the
energy, the domain wall must connect $A$ and $B$ along a straight line.
The energy of such   domain wall is proportional to its length, and,
therefore, the Wilson loop has the area law behavior. The minimal energy
solution is schematically depicted in Fig.~\ref{fig:wall}.

We see that the string tension for the fundamental string is equal to the
tension of the domain wall which separates the two vacua in the theory.
This relation has been discussed a long time ago~\cite{thooft} by 't Hooft.
Parametrically,
this string tension is clearly the same as the adjoint one,
\be
\sigma_f\propto g^2m_{ph}\,,
\ee
although the proportionality constant is different. We will briefly discuss
the relation between the adjoint and the fundamental string tensions
in the next subsection.

Note that
the fundamental string is an absolutely stable topological
object in the $Z_{2}$ invariant theory: the domain wall. It can not break,
if one makes the distance between the two charges larger. In the
effective
theory it is also
obvious since there are no point-like (particle-like)
objects in the theory on which the domain wall can
terminate since there are no dynamical objects with half-integer
winding number.

We interpreted this calculation as the calculation of the potential between
the fundamental adjoint charges -- the time-like Wilson loop. However, in
the Euclidean formulation there is no difference between
the  time-like and
space-like
Wilson loops. Interpreted in this way the calculation becomes a technical
illustration to the argument given in the previous subsection:
space-like Wilson
loops have the area law if the magnetic $Z_2$
symmetry is spontaneously broken.

\subsection{Gluodynamics}

In the weakly interacting case the
effective low-energy Lagrangian can be derived as explained
above in perturbation theory,  combined with the dilute
 monopole gas approximation.
The more interesting regime is certainly that of strong coupling,
which is essentially   pure Yang-Mills theory.
The luxury one has in 2+1 dimensions is that
the weak and  strong coupling regimes are
not separated by a phase transition.\cite{fradkin}
This means that whatever global symmetries the theory has, their
realization must be
the same in the weakly coupled and   strongly coupled vacua.
The existence of the $Z_{N}$ symmetry  is an exact
statement   not related to the weak coupling limit. Therefore, it is
natural to expect that this symmetry must be nontrivially
represented in the effective low-energy Lagrangian.
It is plausible then that the low-energy dynamics at strong coupling is
described by
the same effective Lagrangian which encodes
the spontaneous  magnetic
symmetry breaking.  Certainly, the values of the coupling constants will be
different in   two regimes, but the qualitative behavior should be similar.

Strictly speaking, this is an
assumption rather than   theorem. That is where the difference between
continuous and
discrete symmetries comes in.
Were magnetic symmetry  continuous
(like in the Abelian case), its spontaneous breaking would unambiguously
determine the structure of the effective Lagrangian,
whether  the theory was
weakly or strongly coupled.
With discrete symmetries this is not necessarily the case.
 It could
happen that even though the symmetry is broken, the ``pseudo-Goldstone"
particle is so heavy that it decouples from the low energy dynamics.
For this to happen, though, the symmetry breaking would have to occur at
a very high energy scale.
In gluodynamics this is very unlikely, since the theory has only one
dynamical scale.
In fact, as was discussed in the beginning of this section, the fundamental
string tension determines the scale at which $Z_N$ is broken.
Wilson loops of linear size $l\le(\sigma)^{-1/2}$ do not distinguish between
confining and nonconfining behavior, and hence between the broken
and the unbroken
$Z_N$. The scale of $Z_N$ breaking   is $(\sigma)^{1/2}$
which is precisely a natural dynamical scale of QCD.

Therefore, generally speaking,  we expect that the
pseudo-Goldstone meson stays among the low-energy
excitations not only in the weakly
coupled
limit but also in pure gluodynamics.
If this is the case
the degrees of freedom that enter
the effective Lagrangian in the weakly coupled phase also
interpolate actual low-energy physical states of the
strong coupling regime. That is to say, the radial and phase
components of the vortex field $V$ must correspond to the lightest glueballs
of pure  SU$(N)$ Yang Mills theory.
We can check whether this is the case by
considering the lattice gauge theory data~\cite{teper} on the spectrum.
The radial part of $V$ is
obviously a scalar and has quantum numbers $0^{++}$. The quantum numbers of
the phase are easily determined from the definition (\ref{VQCD}). Those
are $0^{--}$. The spectrum of pure SU$(N)$ Yang Mills theory in 2+1
dimensions was extensively studied recently~\cite{teper} on the  lattice. The
two lightest glueballs for any $N$ are found to have exactly these quantum
numbers. The lightest excitation is   scalar, while the next one is a
charge-conjugation-odd pseudo-scalar
meson with the ratio of the masses roughly~\footnote{Actually
this pattern is firmly established only for $N>2$. At $N=2$
the mass of the pseudo-scalar meson was not   calculated.\cite{teper} The
reason is that it is not clear how to construct a charge-conjugation-odd
operator in  pure gauge SU(2)  lattice theory. So, it is possible that the
situation at $N=2$ is non-generic in this respect. In this case
our strong coupling picture should apply at $N>2$.}
$m_{p}/m_{s}=1.5$ for any $N$.

The situation,
therefore,  is  likely the following. The low-energy physics of
the  SU(2)  gauge theory is always described by the effective
Lagrangian
(\ref{ldualgg}). In the weak coupling regime the parameters are given in
Eqs.~(\ref{couplings1}) and (\ref{couplings2}). Here the
pseudo-scalar particle
is the lightest state in the spectrum
while the
scalar is the first excitation. The pseudo-scalar meson
presents an  ``almost massless photon."  The scalar  meson is the massive Higgs
particle. Moving towards the strong coupling regime (decreasing the Higgs VEV)
one  increases the pseudo-scalar
mass while reducing the scalar mass; the parameters of the effective
Lagrangian change accordingly. The crossover between the weak and   strong
coupling regimes occurs roughly where the scalar and the pseudo-scalar become
degenerate. At strong coupling the degrees
of freedom in the effective Lagrangian are the two lightest glueballs. They
are   still collected in one complex field which nontrivially represents
the exact $Z_{2}$ symmetry of the theory (or $Z_N$ for
 SU(N)).

Of course the spectrum
of pure Yang-Mills theory, apart from the scalar and the pseudo-scalar
glueballs,
contains many other massive glueball states and those are not separated by a
large gap from the two lowest ones. Application of the effective Lagrangian
in the strong coupling regime   has to be considered as qualitative
rather than quantitative. At this qualitative level though,  linear
confinement is an immediate property  of the  effective Lagrangians of
this type.

The fact that the masses of
the scalar and
 pseudo-scalar particles are interchanged in gluodynamics relative to the
weakly coupled regime leads~\cite{kovner3} to some interesting qualitative
differences.
In particular, the structure of the confining string and the interaction
between the strings differ in some important ways.
Let me briefly discuss this aspect.

\begin{figure}
\hskip 2cm
\psfig{figure=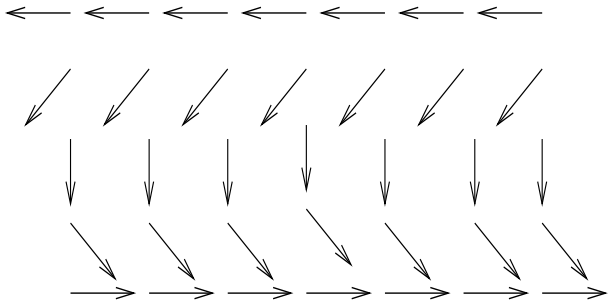,height=1.5in,width=3.2in}
\caption{The structure of the string (domain wall) in the regime when
the pseudo scalar is lighter than  scalar, $m<M$.}
\label{fig:weakwall}
\end{figure}

In the weakly coupled regime the phase of $V$ is
much lighter than the radial part. A cartoon of the fundamental string in
this situation is depicted in Fig.~\ref{fig:weakwall}.

The radial part $\rho$ being very heavy practically does not change inside
the string. The value of $\rho$ in the middle of the string can
be estimated from the following simple argument. The width of
the region where
$\rho$ varies from its vacuum value $\mu$ to the value $\rho_0$ in the middle,
is of the order of the inverse mass of $\rho$. In terms of the energy per
unit length
 this variation costs
\begin{equation}
\sigma_\rho\sim M(\mu-\rho_0)^2+x{\frac{m^2}{M}}\rho_0^2\,,
\end{equation}
where $x$ is a dimensionless number of order unity.
The first term is the contribution of the kinetic term of $\rho$ while
the second
contribution comes from the interaction term between $\rho$ and $\chi$ due
to the fact that the value of $\chi$ in the middle of the string differs
from its vacuum value.
Our notations are such that $m$ is the mass of the pseudo-scalar
particle and $M$ is
the mass of the scalar one.
Minimizing this with respect to $\rho_0$ we find
\begin{equation}
\rho_0=\mu (1-x{\frac{m^2}{M^2}})\,.
\end{equation}
Thus, even in the middle of the string the difference in the
value of $\rho$ and its VEV is of the second order in the small ratio $m/M$.
Correspondingly, the contribution of the energy density of $\rho$ to the
total energy density is also very small,
\begin{equation}
\sigma_\rho\sim {\frac{m }{M}}m\mu^2\,.
\end{equation}
This is to be compared with the total tension of the string which is
contributed mainly by the pseudo-scalar phase $\chi$,
\begin{equation}
\sigma_\chi\sim m\mu^2\,.
\end{equation}
We obtain this by estimating the kinetic energy of $\chi$ on a
configuration of width $1/m$ where $\chi$ changes~\footnote{The fact
that the heavy radial field $\rho $ practically does not
contribute to
the string tension is natural from the point of view of
decoupling. In the limit of infinite mass $\rho $ should decouple from the
theory without changing its physical properties. It is  very
different,however, from the situation in superconductors. In the
superconductor of the
second kind, where the order parameter field is much heavier than the
photon ( $\kappa >
\frac{1}{\sqrt{2}}$) the magnetic field and the order parameter give
contributions of the same order
(up to logarithmic corrections $O(\log \kappa) $)
to the energy of the Abrikosov
vortex. This is the consequence of the fact
that the order parameter itself is forced to vanish in the core of the
vortex, and therefore even though it is heavy, its variation inside
the vortex is large. An even more spectacular situation arises if we
consider a domain wall between two vacuum states in which the heavy field
has different values.\cite{kovner4} In this situation the contribution of
the heavy field $\phi$ to the tension would be
\begin{equation}
\sigma _{heavy}=M(\Delta \phi)^{2}
\end{equation}
where $\Delta \phi$ is the difference in the values of $\phi$
on both sides of the
wall. For
a fixed $\Delta \phi$ the energy density diverges when $\phi$
becomes
heavy. In the present case this does not happen since the two vacua which are
separated by the domain wall differ only in VEV of the light field $\chi $
and not the heavy field $\rho $.} by an amount of order $1$.

\begin{figure}
\hskip 2.5cm
\psfig{figure=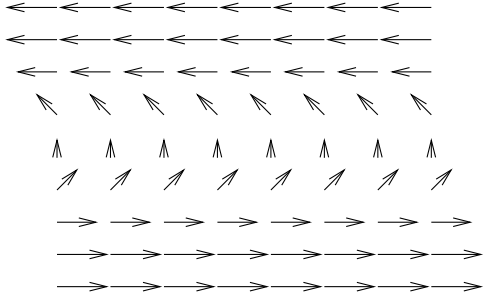,height=1.5in}
\caption{The structure of the string (domain wall) in the regime when
the pseudo-scalar is heavier than the scalar, $m>M$.}
\label{fig:strongwall}
\end{figure}

Let us now consider the domain wall (or fundamental string) in the
opposite regime,
when the mass of the scalar meson is much smaller than that of the
pseudo-scalar one.  Now, the profile of the fields in the wall  is very
different.
A cartoon of this situation is given in Fig.~\ref{fig:strongwall}.

We will use the same notations, denoting the mass of
the pseudo-scalar by $m$ and the mass of the scalar by $M$, but now $m\gg M$.
Let us again estimate the string tension and the contributions of the scalar
and  of a pseudo-scalar to it. The width of the region in which the
variation
of $\rho $ takes place is of the order of its inverse mass. An estimate of the
energy density of the $\rho $ field is given by the contribution of the
kinetic term
\begin{equation}
\sigma _{\rho }\sim M(\mu -\rho _{0})^{2}\,.
\end{equation}
The width of the region in which the phase $\chi $ varies is $\sim 1/m$. In
this narrow strip the radial field $\rho $ is practically constant and is
equal to $\rho _{0}$. Hence, the kinetic energy of $\chi $   contributes
\begin{equation}
\sigma _{\chi }\sim m\rho _{0}^{2}\,.
\end{equation}
Minimizing the sum of the above two contributions with respect to
$\rho _{0}$ we
find
\begin{equation}
\rho _{0}\sim {\frac{M}{m}}\mu \ll\mu \,.
\end{equation}
Also
\begin{eqnarray}
\sigma _{\chi }  \sim  {\frac{M}{m}}M\mu ^{2} \,, \qquad
\sigma _{F} =\sigma _{\rho }\sim M\mu ^{2}\,.
\end{eqnarray}

Now the radial field is very small in the core of the string.
The energy density is contributed almost entirely by the scalar rather
than by the
pseudo-scalar field.\footnote{Again, this is in agreement with
decoupling.
The heavier
field does not contribute to the energy, even though its values
on the opposite
sides of the wall differ by $O(1)$. Its contribution to the energy
is suppressed by the factor $\rho _{0}^{2}$ which is very small inside
the wall.}

The extreme situation $m\gg M$  is not
realized in the non-Abelian gauge theory. From the lattice simulations we
know that in actuality even in the case of pure Yang-Mills  the ratio
between the
pseudo-scalar and scalar masses is about $1.5$ -- not a very large number. The
analysis of the previous paragraph   does not reflect the situation
in the strongly coupled regime of the theory. Rather we expect that the
actual profile of the string is somewhere in between
Fig.~\ref{fig:weakwall}
and Fig.~\ref{fig:strongwall},
although somewhat closer to Fig.~\ref{fig:strongwall}.
The widths of the string in terms of the
scalar and pseudo-scalar fields are of the same order, although the scalar
component is somewhat wider. The same is applicable to the contribution to the
string tension. Both glueballs contribute, with the scalar contribution being
somewhat larger.\footnote{One
also expects similar non-negligible contributions
from higher mass glueballs which are not taken into
account in our effective Lagrangian framework.}

The interaction between two
domain walls (confining fundamental strings) in the two extreme
regimes is also quite different. In the weakly coupled region we
can disregard the variation of $\rho $. For two widely separated parallel
strings the
interaction energy comes from the kinetic term of $\chi $. This obviously
leads to a repulsion, since for both strings in the interaction region the
derivative of the phase is positive. On the other hand, if the pseudo-scalar
is very heavy the main interaction at large separation is through the
``exchange'' of the scalar. This interaction is clearly attractive, since if
the strings overlap, the region of
the space where $\rho $ is different from its
value in the vacuum is reduced relative to the situation when the strings
are far apart.

Therefore, the situation is   very similar to that in superconductivity. The
confining strings in the weakly coupled and strongly coupled regimes behave
like the Abrikosov vortices in the superconductor of the second and first kind,
respectively. This observation has an immediate implication for the string
tension of the adjoint string. As we have discussed in the previous section
the phase $\chi $ changes from $0$ to $2\pi $ inside the adjoint string. The
adjoint string   can be pictured as two fundamental strings running
parallel to each other. In the weak coupling regime the two
fundamental
strings
repel each other. Therefore, two fundamental strings within the adjoint string
 will not overlap, and the energy of the adjoint string is twice
the energy of the fundamental one,
\begin{equation}
\sigma _{Adj}=2\sigma _{F}\,.
\end{equation}

In the strongly coupled region the situation is quite different. The strings
attract. It is clear that the contribution of $\rho $ to the energy will be
minimized if they overlap completely. In this case the contribution
of $\rho $ in the fundamental and adjoint strings will be roughly the same.
There will still be a repulsion between the pseudo-scalar cores of the two
fundamental strings, so presumably the core energy will be doubled. In this
way we
arrive at an estimate
\begin{equation}
\sigma_{\rm Adj}=\sigma _{F}+O({\frac{M}{m}}\sigma _{F})\,.
  \label{sst}
\end{equation}

Again the situation is more complicated in the limit
of pure Yang-Mills. The
scalar is lighter, therefore the interaction at large distances is
attractive. However, the pseudo-scalar core size and its contribution to the
tension are not small. In other words, $M/m$ in Eq.~(\ref{sst}) is a number
of order one.

Thus, in broad terms  the relation between the weakly and strongly coupled
confining theories is similar to the relation of the superconductors of
the second and first kind.

 \section{Fundamental Matter.}

The discussion in the preceding sections pertained only to theories without matter fields in
fundamental representation. Obviously it is very important to understand what becomes of
the magnetic symmetry and the associated order parameter if fundamental matter is present.
In this section we address this question\cite{fosco}.

For
simplicity we will consider scalar quarks. The Lagrangian of the
theory we are interested in is
\begin{equation}
L=-{1\over 4}{\rm Tr}F^2+|D_\mu\Phi|^2-M^2\Phi^*\Phi
\label{lqcd}
\end{equation}
where the scalar field $\Phi^\alpha$ transforms according to the
fundamental representation of the $SU(N)$ color group.
\subsection{$Z_N$ as a local symmetry.}
First thing to note is that the fundamental Wilson loop still
commutes with the Hamiltonian. This is obvious, since the extra
terms in the Hamiltonian in the presence of the quarks do not
involve electric field operator, but only vector potential. The
Wilson loop commutes with the vector potential, and thus with the
additional terms in the Hamiltonian.  The theory thus still has
the magnetic $Z_N$ symmetry.

This may seem somewhat surprising at first sight.  We have shown
above that spontaneous breaking of the magnetic $Z_N$ implies the
area law for the Wilson loop, and conversely the perimeter law of
$W$ implies unbroken $Z_N$. In the theory with fundamental charges
the Wilson loop is known to have perimeter law due to breaking of
the confining string at any finite value of the fundamental mass
$M$.  We might then conclude that the magnetic $Z_N$ is restored
at any, arbitrarily large value of $M$ but is broken at
$M\to\infty$. The common lore is that the $Z_N$ breaking phase
transition for $N>2$ is first order, and this then implies a
discontinuous behaviour of the theory in the infinite mass limit.
This of course is completely counterintuitive and in fact plain
wrong.  The caveat in this line of reasoning is the following. The
relation between the behaviour of the Wilson loop and the mode of
the realization of the magnetic symmetry hinges crucially on the
existence of a local order parameter of the magnetic $Z_N$. In the
absence of such an order parameter it is not true that the Wilson
loop locally changes the quantum state inside the loop and this
invalidates the whole argument. In particular in the absence of a
local order parameter, the $Z_N$ symmetry can be spontaneously
broken but the Wilson loop can have a perimeter law.

In fact it is easy to see that for any finite $M$ the magnetic $Z_N$
does not have a local order parameter. The only candidate for such an
order parameter is the vortex operator $V(x)$ defined in
eq.(\ref{v2}), since it has to be local also relative to the purely
gluonic operators.
However in the presence of fundamental quarks the operator $V(x)$ is
not local anymore.  To see this consider the
dependence of $V_C(x)$ on the curve $C$ which enters its definition.
As before the operators $V_C$ and $V_{C'}$ are related by
\begin{equation}
V_C(x)=V_{C'}(x)\exp\{{4\pi i\over gN}\int_{S}d^2x {\rm Tr}\partial_iYE^i\}
\end{equation}
where $S$ is the area bounded by $C-C'$. As before, due to the Gauss'
law the integral in the exponential is equal to the total hypercharge
in the area $S$.  However the hypercharge of fundamental quarks has
eigenvalues $\pm g/2$. The extra phase factor is therefore not unity
anymore but can rather take values $\exp\{2\pi i/N\}$ depending on the
state and the choice of the contour $C$.

The status of the magnetic $Z_N$ is thus quite different in a theory
with fundamental quarks - it does not have a local order parameter.
Nevertheless it is clear that at least as long as the mass $M$ is
large $M/g^2>1$, the relevant degrees of freedom for the effective
infrared dynamics should still be the vortices $V$ and the main factor
which determines their dynamics should still be the magnetic $Z_N$. At
large $M$ the dependence of $V$ on the curve $C$ is weak, since the
probability of the vacuum fluctuations which involve fundamental
charges is small.  The probability of appearance of a virtual $q\bar
q$ pair separated by a distance $l$ is suppressed by the exponential
factor $\exp\{-Ml\}$.  The typical distance scale for the "glueball"
physics is $1/g^2$.  Thus at these distances such fluctuations are
unimportant and should not affect much the dynamics.  Things are
different however if one is interested also in the baryonic sector of
the theory. The baryons are necessarily heavy and in order to be able
to discuss their structure we must understand the main dynamical
effects also at shorter distances.

Therefore our aim now is to understand
what is the main effect of the non locality discussed above on the
dynamics of magnetic vortices.

The situation we have just described - a symmetry without a local
order parameter - is not exceptional in quantum field theory. This is
precisely the property of the global part of any Abelian gauge group.
Consider for example quantum electrodynamics. The global electric
charge is of course a physical gauge invariant charge with the
corresponding gauge invariant local charge density.
\begin{equation}
Q=\int d^2x\rho
\end{equation}
Nevertheless there is no local operator that carries this charge. This
is a direct consequence of the Gauss' law
\begin{equation}
\partial_iE_i=g\rho
\end{equation}
Any physical, gauge invariant operator that carries $Q$ must also
carry the long range electric field, which can not fall off faster
than a power of the distance.  The gauge invariant QED Lagrangian is
written in terms of "local" charged fields $\phi$. But appearances are
deceptive: these fields are not gauge invariant, and therefore not
physical. A gauge invariant charged field can be constructed from
$\phi$ by multiplying it by a phase factor
\begin{equation}
\phi_{phys}(x)=\phi\exp\{ig\int d^2y e_i(x-y)A_i(y)\}
\end{equation}
with the $c$-number field $e_i$ satisfying
\begin{equation}
\partial_ie_i=\delta^2(x)
\end{equation}
For any $e_i$ satisfying this condition the operator $\phi_{phys}$ is
gauge invariant, and therefore physical. It is however necessarily
nonlocal. Different choices of $e_i$ define different gauge invariant
operators and correspond to different gauge fixings. Thus for
$e_i(x)=x_i/x^2$ the field $\phi_{phys}$ is the field $\phi$ in the
Coulomb gauge, while $e_i(x)=\delta_{i1}\delta(x_2)\theta(x_1)$
corresponds to the axial gauge $A_1=0$, and so on.  Different
definitions of $\phi_{phys}$ differ from each other by a phase factor,
which is precisely the gauge ambiguity of the original field $\phi$.

The $U(1)$ gauge group is the most natural Abelian gauge symmetry
to consider in continuum field theory. One can however also
consider discrete groups like $Z_N$\cite{ZN}.  In this case again
no local operator that carries the global $Z_N$ charge exists. The
various gauge invariant charged operators are nonlocal and differ
from each other by a local $Z_N$ - valued phase. These different
operators again correspond to different gauge fixings of the local
$Z_N$ group.

This is precisely the structure that emerged in our discussion in the
earlier part of this section.  We have

1. Global magnetic $Z_N$ symmetry generated by the fundamental Wilson
loop.

2. The set of nonlocal vortex operators $V_C(x)$, which all carry the
$Z_N$ charge and differ from each other by a $Z_N$ valued phase
factors.

It is very suggestive therefore to think about $V_C(x)$ as of
different gauge fixed versions of a field charged under local $Z_N$.
This leads us to expect that the low energy theory we are after should
be a $Z_N$ gauge theory of the magnetic vortex field $V$.

In fact coming back to the discussion in the beginning of this
section, we see that from this vantage point it is obvious why the
Wilson loop has a perimeter law, even if the global $Z_N$ is broken
spontaneously. The action of a Wilson loop of a finite size inside the
contour is a gauge transformation. Thus in physical terms locally
inside the contour the new state is {\it the same\/} as the old one and
so the overlap between the two locally is unity. The only nontrivial
contributions to the overlap come from the region close to the contour,
thus giving the perimeter law.

We will now construct the effective low energy locally $Z_N$
invariant theory, and discuss its relation to the original QCD
Lagrangian. Although it is possible to  give a more `analytic'
derivation of the effective Lagrangian, we prefer to write it down
directly guided by the previous discussion. We will then explain
the physical meaning of the various fields that appear in it.

\subsection{Gauging the Wilson loop.}
The easiest way to construct a $Z_N$ gauge theory in the continuum
is to consider a $U(1)$ gauge theory with the Higgs field of
charge $N$ which has a large expectation value~\cite{ZN}.
Consider therefore the following Lagrangian:
\begin{eqnarray}
\label{efflfund} L&=&-{1\over 4e^2}f_{\mu\nu}^2+|(\partial_\mu-i{1\over
N}b_\mu)V|^2+ |(\partial_\mu-ib_\mu)U|^2 \nonumber\\
&-&\lambda(V^*V-\mu^2)^2 - \xi(V^NU^* +V^{*N}U) -
\tilde\lambda(U^*U-u^2)^2 \;.
\end{eqnarray}
Here $f_{\mu\nu}=\partial_\mu b_\nu-\partial_\nu b_\mu$. We take the
parameters such that, $\tilde\lambda>>\lambda$ and $u^2>>\mu^2$.

The `Higgs' field $U$ has a large expectation value and breaks the
$U(1)$ gauge symmetry down to its $Z_N$ subgroup
$V(x)\to\exp\{i{2\pi n(x)\over N}\}V(x)$.
Below the scale
determined by the expectation value $u$, the field $U$ is
practically frozen and its fluctuations are unimportant. In this
regime the model indeed describes the locally $Z_N$ invariant
theory. The global part of the gauge group is our $Z_N$ magnetic
symmetry generated by the Wilson loop.  The larger $U(1)$ gauge
structure at this point is just an auxiliary trick which enables
us to write down a discrete gauge theory in continuous notations.
We will however see later that it does in fact has a real physical
meaning of its own and arises naturally in the effective theory.

Let us first discuss how this Lagrangian reduces to the effective
Lagrangian of the pure Yang-Mills theory eq.(\ref{ldualgg}) in the
limit of the large quark mass.  In this limit not only the field $U$
must decouple, but also the gauge interactions of the field $V$ must
vanish.  There is another consistency requirement. In the limit of
zero gauge coupling $e^2\to 0$ the gauge $U(1)$ symmetry
becomes global $U(1)$ and is broken due to non vanishing expectation
value of $U$. The spectrum therefore contains a massless Goldstone
boson. This Goldstone boson is of course the longitudinal component of
the gauge field $b_\mu$. There is however no such massless particle in
the pure gluodynamics nor in the effective Lagrangian
eq.(\ref{ldualgg}).  This means that the couplings in
eq.(\ref{efflfund}) should depend on the quark mass in such a way that
the vector particles remains heavy for any finite $M$ and its mass
goes to zero very sharply only in the limit when it is completely
decoupled.  In terms of the Goldstone boson couplings it means that
$f_\pi$ must be larger than any scale relevant to the dynamics of the
vortex field $V$.  All these conditions can be met by choosing for
example
\begin{equation}
e^2={y\over M}, \ \ \ \ \ \ \ u^2=xM,\ \ \ \ \ \ \ \ \ \xi={\zeta\over u}
\label{largem}
\end{equation}
With this choice the mass of the vector boson $m^2=e^2u^2$ stays
finite as $M\to \infty$ and can be arbitrarily large. The
Goldstone boson in the decoupling limit has an infinite $f_\pi$ and is
completely decoupled just as the "invisible axion".

How do the basic fields present in the
effective Lagrangian arise in the fundamental theory eq.(\ref{lqcd})?
To understand this, let us start with considering the symmetries of the theory.

The Lagrangian eq.(\ref{efflfund}) has two global $U(1)$
symmetries. One is the global part of the local $U(1)$ with the
conserved current $ej^T_\mu=\partial_\nu f_{\nu\mu}$.  The other
current, conserved by virtue of the homogeneous Maxwell equation
is $\tilde f_\mu=\epsilon_{\mu\nu\lambda}f_{\nu\lambda}$.  The two
charges have quite different nature. The second one, the dual
magnetic flux
\begin{equation}
\Phi_D=\int d^2 x \tilde f_0
\end{equation}
has a local order parameter. It can be constructed in a way similar to
the vortex field in QED \cite{kovner},\cite{kovner1}. The global $U(1)$
gauge charge, on the other hand does not have a local order parameter,
as discussed earlier.

Both these conserved currents also exist in the fundamental
theory. It is fairly straightforward to identify them.  The QCD
Lagrangian has one obvious global $U(1)$ charge - the baryon
number. This charge has local order parameters - gauge invariant
baryon fields of QCD, and is therefore identified with the dual
magnetic flux
\begin{equation}
{1\over 2\pi}\tilde f_\mu=J^B_\mu,\ \ \ \ \ \ Q_B=\Phi_D
\label{f}
\end{equation}

The second charge can be expressed in terms of the spatial current
components of the first one
\begin{equation}
Q^T=\int d^2xj^T_0=\int d^2x \partial_i\Big[{1\over e^2}\epsilon_{ij}\tilde f_j\Big]
\label{QT}
\end{equation}
It is thus the vorticity associated with the baryon number current.

The vortex operator $V$ which appears in eq.(\ref{efflfund}) is
not gauge invariant and is only physical after complete gauge
fixing of the $U(1)$ gauge group. After such a gauge fixing (which
amounts to multiplying $V$ by an operator valued phase) the Gauss
law requires that on the physical states the physical operator $V$
carry the charge $Q_T$.  Due to the identification
eqs.(\ref{f},\ref{QT}) this leads to an interesting conclusion
that any physical operator $V$ in the effective theory creates a
vortex of the original baryon number current.

This somewhat unexpected conclusion is in fact quite natural for
an eigenoperator of the magnetic $Z_N$. Consider the vortex
operator $V$ defined by eq.(\ref{v2}). In QCD just like in pure
gluodynamics, it has an alternative representation of an operator
of the singular gauge transformation of the form eq.(\ref{VQCD})
\begin{equation}
V_C(x)=\exp\{{2i\over gN}\int d^y \epsilon_{ij}{x_i-y_j\over (x-y)^2}{\rm Tr}(YE_j(y))+
\theta(x-y)J_0^Y(y)\}
\label{v3}
\end{equation}
The only difference is that now $J_0^Y$ is the hypercharge operator
due to both, gluons and fundamental quarks
\begin{equation}
J^Y_0=ig\Big[{\rm Tr}Y[A_i, E_i]+Y_{\alpha\beta}(\Phi^*_\alpha\Pi_\beta-
\Phi_\alpha\Pi^*_\beta)\Big]
\end{equation}
Consider the action of this operator on a quark field $\Phi$. The
transformed quark field $\Phi'=V^\dagger_C\Phi V_C$ is a gauge
transform of $\Phi$ everywhere except along the branch cut of the
function $\theta$. Across this cut the phase of $\Phi'$ is
discontinuous - it jumps by $2\pi/N$ for all color components of
$\Phi'$.  Due to this discontinuity the baryon current - the global
$U(1)$ current of $\Phi$ - does not vanish at points along the cut.
Calculating explicitly the action of $V$ on the baryon number current
we find
\begin{equation}
V^\dagger_CJ^B_i(x)V_C=iV^\dagger_C(\Phi^*\partial_i\Phi-\partial_i\Phi^*\Phi)V_C=
J^B_i(x)+{2\pi\over N}n_i^C(x)\delta(x\in C)\Phi^*\Phi(x)
\end{equation}
where $n_i^C(x)$ is a unit vector normal to the branch cut $C$ at the
point $x$.

It is natural to define the local vorticity associated with the
baryon number as
\begin{equation}
\rho_T=i\epsilon_{ij}\partial_i[{(\Phi^*\partial_j\Phi-\partial_j\Phi^*\Phi)\over
\Phi^*\Phi}]
\label{vort}
\end{equation}
The vortex operator $V$ therefore creates a vortex of baryon number
current with fractional vorticity $2\pi/N$
\begin{equation}
V^\dagger_C(x)\rho_T(y)V_C(x)=\rho_T(y)+{2\pi\over N}\delta^2(x-y)
\end{equation}

The operator $U$ due to the Gauss' law also carries baryon vorticity.
Since its gauge coupling is $N$ times the coupling of $V$, it creates
one unit of vorticity.

This simple exercise also helps us to identify the value of
the gauge coupling constant $e^2$ in the effective theory.  Comparing
eq.(\ref{vort}) with eq.(\ref{QT}) we find
\begin{equation}
e^2\propto \Phi^*\Phi
\label{e2}
\end{equation}
The same relation is obtained by comparing the current algebra in the
fundamental and the effective theories.  The commutator of the baryon
charge density with the baryon current density in the fundamental
theory is
\begin{equation}
[J^B_0(x),J^B_i(y)]=i\Phi^*\Phi\partial_i\delta^2(x-y)
\end{equation}
In the effective theory using the canonical commutators that follow
from eq.(\ref{efflfund}) and the identification eq.(\ref{f}) we find
\begin{equation}
[J^B_0(x),J^B_i(y)]=i{e^2\over 4\pi^2}\partial_i\delta^2(x-y)
\end{equation}
Again we deduce eq.(\ref{e2}).  The operator on the right hand
side of this equation in the effective theory is indeed a
constant. The effective theory should be valid at long distances.
In this regime in the leading order in the derivative expansion
the operator $\Phi^*\Phi$ should be approximated by its
expectation value. Taking into account fluctuations of
$\Phi^*\Phi$ is tantamount to including higher derivative terms in
the effective Lagrangian eq.(\ref{efflfund}).  At this, higher
order in derivative expansion the gauge coupling constant would
become a dynamical field. While this is perfectly legitimate, it
is beyond our present framework.

When calculating the expectation value of $\Phi^*\Phi$ we should
remember that it has to be calculated with the cutoff $\Lambda$
appropriate for the effective theory.  This cutoff must be above
the characteristic scale of the pure gluodynamics (determined by
the string tension) but below the heavy quark mass.  With this in
mind we get
\begin{equation}
<\Phi^*\Phi>_\Lambda=\int_0^{\Lambda} {d^2p\over 8\pi^2}{1\over (p^2+M^2)^{1/2}}\propto
{\Lambda^2\over M}
\end{equation}
So that finally
\begin{equation}
e^2\propto{\Lambda^2\over M}
\end{equation}
which is consistent with the expected scaling in the large mass limit
eq.(\ref{largem}).

\subsection{The baryon and the bag.}\label{baryon}
Having understood the origin of the fields and the symmetries in
the effective theory, we would like to see how it encodes the
qualitative features of the low energy QCD physics.  Since we are
considering a heavy quark theory, below the fundamental mass scale
the spectrum should be the same as in pure gluodynamics.  We have
already seen that, in the infinite mass limit, the effective
theory reduces to that of (\ref{ldualgg}). Indeed, even at finite
but large $M$, this is the case at low energies. Since the VEV of
the field $U$ is large, we can impose the unitary gauge condition
on it. In this unitary gauge the phase of $U$ disappears. The
modulus of $U$ is very heavy, and so is the vector field $b_\mu$.
Thus at low energies we recover the effective theory of pure glue
sector.  There is however a set of configurations, on which the
unitary gauge can not be imposed. Those are configurations in
which $U$ vanishes at some points in space.  Indeed it is these
configurations that are important for the baryonic sector.  Recall
that $U$ is a vortex of baryon number current. Thus, by duality
one expects that the baryon charge is associated with the vortex
configuration of the field $U$. In the core of the vortex the
field $U$ has to vanish, and so the unitary gauge is not
admissible. Thus in the baryon sector we can not think of $U$ as
frozen at its expectation value and instead have to treat it as a
dynamical field.

That the baryon does indeed carry vorticity of the field $U$, can
be seen by the following simple argument.  The baryon number is
represented in the effective theory by the dual magnetic flux. The
baryonic state must therefore be the dual magnetic vortex.  Such a
vortex in a nonsingular gauge has a vector potential of the form
$b_i=\epsilon_{ij}{x_j\over x^2}$.  To have a finite energy it
must be accompanied by the winding of the phases of both $V$ and
$U$. Since $U$ carries $N$ times the charge of $V$, the only
states that are allowed energetically are those that carry $N$
vortices of $U$.

This is natural in view of the `dual' relation between the
effective and the fundamental theories. The field $U$ is the
`vortex' dual to the fundamental quark.  We thus expect that its
elementary vortex would represent the fundamental quark itself,
and so finite energy states must contain $N$ such elementary
vortices.  A single vortex must be confined.  In the same sense,
$V$ is `dual' to the adjoint gluon field.  The elementary vortex
of $V$ is then, in a sense, the `constituent' gluon.  In fact,
such single gluon should not exist as a finite energy state
either, and we expect it also to be confined. The single vortex of
$V$ should therefore bind either with the anti-vortex of the same
type, or with $N$ vortices of $U$.  The former type of state is a
glueball, and exists in the pure glue theory~\cite{kovner3}, while
the latter type is a baryon.

Interestingly enough, this line of reasoning leads us to expect
that the baryon must have a bag-like structure. Namely, the quarks
are bound to the $V$ field vortex. Inside this vortex, the value
of $V$ is small - it vanishes in the middle, and than rises quite
slowly (relative to the scale of $1/M$) towards the edges. Recall
that $V=0$ corresponds to a non confining state~\cite{kovner3}.
The quarks are therefore sitting in the `perturbative' region of
space - where there are no confining forces.  Only when they
separate far from each other - into the region with non vanishing
$V$, the linear potential pulls them inside again.

Let us look at this more carefully. Consider for simplicity the $Z_2$
symmetric case $N=2$.  The baryon is the dual vortex with dual
magnetic flux $2\pi/e$.  Far from the vortex core,
the field configuration is pure gauge, with the phases of $V$ and $U$
following the vector potential:
\begin{equation}
b_i=\epsilon_{ij}{x_j\over x^2},\;\;\; V(x)=ve^{i \alpha(x)},\;\;\;
U(x)=ue^{2 i \alpha(x)} \;.
\end{equation}
The parameters of the model are such that the field $V$ is much
lighter than both $U$ and $b_i$. Thus the size of the vortex core
of $V$ is large - of the order of the inverse glueball mass. The
two $U$-vortices which have a very small size core, sit inside
this core. Since the length associated with the dual magnetic
field is much smaller than the core size of the $V$ - vortex, the
dual flux is concentrated on the $U$ - vortices. From the low
energy point of view, the picture is that two point-like magnetic
vortices sit inside a soft core of a $V$ field vortex.  The field
configuration looks roughly as depicted on figure 7.

The magnetic flux is concentrated at the points $A$ and $B$. The phase
of the field $U$ follows the variation of the vector potential very
closely. The phase of $V$ is also trying to do that, but it can not
quite follow it all the way, since on the line between the two
vortices it would have to be discontinuous. The most important energy
contribution (apart form the core energy of small vortices) comes
therefore from the vicinity of this line. The phase of $V$ obviously
has to interpolate across this line between the values $0$ and $\pi$.
Since the modulus of $V$ is not extremely rigid, it will be smaller
along this line than in the immediate neighborhood.  Both, the
variation of the modulus and the phase of $V$ along the line
connecting the small vortices contribute to the energy which is
clearly linear in the distance $|A-B|$.

\begin{figure}
\hskip 4cm
\psfig{figure=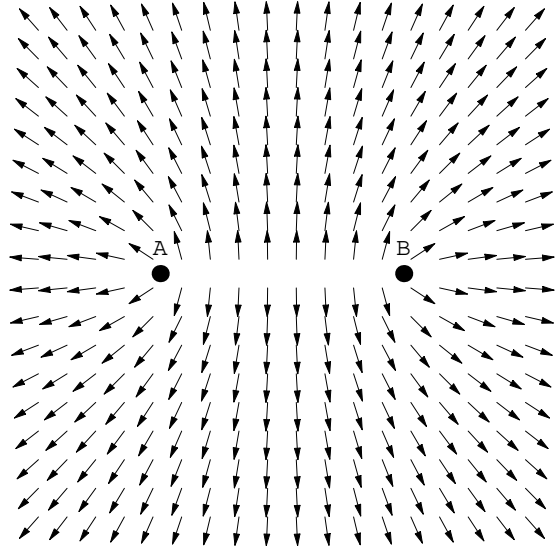,height=3in} \caption{The `baryon'
configuration for $Z_2$. Arrows represent the
  direction (phase) of $V$, and the big dots at $A$ and $B$ correspond
  to the positions of the $U$ field vortices.} \label{fig:bag}
\end{figure}

To study the structure of the baryon in the effective theory more
quantitatively let us fix the gauge such that
$$
 U(x)\,=\,u\, \exp [ i \, \theta (x) \,]
$$
\begin{equation}
 \label{eq:uconf}
\theta (x) \;=\; \arctan (\frac{y}{x-a}) + \arctan(\frac{y}{x+a}) \;.
\end{equation}
This is a valid gauge choice for the configuration with
two vortices of unit winding at the positions $(a,0)$
and $(-a,0)$.

Due to the conditions on the parameters of our model, the vector field
$b_\mu$ follows the phase of $U$ in the whole space. Thus
\begin{equation}
  \label{eq:bconf}
b_0=0,\ \ \ \ \ b_j = - \left[\epsilon_{jk}\frac{(x - x^{(l)})_k}{|x - x^{(l)}|^2}
\,+\, \epsilon_{jk}\frac{(x - x^{(r)})_k}{|x - x^{(r)}|^2} \right]
\end{equation}
where: $x_1 \equiv x$, $x_2 \equiv y$, and $l$, $r$ are the positions
of the left and right vortices, respectively: $x^{(l)} = (-a,0)$,
$x^{(r)} = (a,0)$.

There are two interesting limiting situations. The first is when the
distance $a$ is larger than the dynamical distance scales
$(\lambda\mu^2)^{-1/2}$ and $(\xi u^2)^{-1/2}$ which determine the
masses of the glueballs. The second is the
reverse, that is when the two vortices are sitting well inside the
glueball correlation length. Let us look at them in turn.

When the distance between the vortices is large we expect the
potential between them to be linear with the string tension calculated
in the theory without the dynamical $U$ field.
We will study the interquark potential in our low energy theory in the
semiclassical approximation.

To find the minimal energy configuration, we have to solve the
classical equations of motion for the field $V$ at a fixed
configuration of $U$ and $b_i$ given by
eqs.(\ref{eq:uconf},\ref{eq:bconf}). Let us concentrate on the
points which are close to the $x$ - axis, with $|x|<<a$. The main
contribution to the energy comes from this region of space. In
this region the vector potential $b_i$ vanishes, and the phase of
the field $U$ is zero. Thus the equations of motion for the field
$V$ are the same as in the pure gluodynamics. Also as long as we
stick to this region, the configuration of $V$ is translationally
invariant in the $x$ direction. What determines the energy then
are the boundary conditions on the field $V$. In this
configuration clearly the phase of the field $V$ is $\pi/2$ far
above the $x$ axis and $-\pi/2$ far below the axis, fig.7. Thus
both, the equations and the boundary conditions are precisely the
same as for the domain wall separating the two degenerate vacua in
the effective theory of pure gluodynamics ($e^2=0,\ \
\tilde\lambda\to\infty$). There is an extra contribution to the
energy that comes from the region of space close to the points $A$
and $B$. But this energy does not depend on $a$ and is subleading
for large $a$. The rest of the space does not contribute to the
energy, since the field configuration there is pure gauge.

Thus as expected the energy in this regime is $E=a\sigma$, where
$\sigma$ is the domain wall tension (fundamental string tension)
calculated in pure gluodynamics~\cite{kovner3}.

It is also interesting to consider the opposite situation, when
the distance between the quarks is smaller than the glueball
correlation length. This is the regime in which we do not expect
to see any stringy structure. Instead we can ask whether the
lowest energy configuration has any resemblance to a bag. To study
this question we take the limit $a\to 0$. In this case the phase
of the field $V$ will follow the phase of $U$ in the whole space
\begin{equation}
  \label{eq:imag1}
  (\frac{V}{V^*})^2 \;=\; \frac{U}{U^*} \;.
\end{equation}
Given this condition, only the variation of the radial component of
$V$ has to be determined. Since the problem has rotational symmetry,
the equation of motion for the modulus $\rho$
becomes
\begin{equation}
  \label{eq:eqrho1}
 -\frac{d^2 \rho}{d r^2}\,-\, 2 \eta  \rho
 + 2 \lambda \rho^3\;=\;0\;.
\end{equation}
with  $\eta \;=\; \lambda \mu^2 - \xi u$, which is assumed to be
positive throughout this paper. The relevant boundary condition in this
case is that at infinity $\rho$ approaches its vacuum value
$\rho_{r\to\infty}\to v$ while in the vortex core it
vanishes $\rho(0)=0$. With these boundary conditions, eq.
(\ref{eq:eqrho1}) has the familiar form of the $\varphi^4$ static kink
equation in $1+1$ dimensions with the solution
 \begin{equation}
  \label{eq:rhor}
  \rho (r) \;=\; v \;
  \frac{1 \,-\, e^{-2 \sqrt{\eta} r}}{1 \,+\, e^{-2 \sqrt{\eta} r}} \;.
\end{equation}
The energy of this solution is
\begin{equation}
E[V]_{a=0} \;=\; \lambda \int_0^\infty \, 2 \pi dr r \,
(v^4 - \rho^4) \;=\; \frac{\pi^3}{12} \, \frac{\eta}{\lambda}
\;=\; \frac{\pi^3}{12} \, ( \mu^2 - \frac{\xi u}{\lambda} )\;.
\end{equation}

The picture is thus just as described in the previous subsection. Since
the two quark state is accompanied by the vortex of the field $V$, the
quarks effectively `dig a hole' in the vacuum. In their immediate
vicinity the modulus $\rho$ vanishes, and therefore there is a `bag' of
the nonconfining state. The radius of this bag is given by the mass of
the scalar glueball $2 \sqrt{\eta}$.

It is interesting to note that, although for large separation $a$ the
energy of the string gets contributions from both, the scalar and the
pseudoscalar glueballs (the modulus and the phase of $V$), the `bag
constant' is determined solely by the scalar glueball. For small $a$
the phase of $V$ is not excited and only $\rho$ deviates from the
vacuum state inside the `bag'. This is consistent with the common lore
that the inside of the bag is distinguished from the outside by the
value of the $F^2$ condensate. In fact since $\rho$ has vacuum quantum
numbers and interpolates in our effective theory the scalar glueball,
it is naturally associated with the operator $F^2$ which has a large
overlap with the scalar glueball in QCD.

The bag we are talking about here arises in a very different
situation than in the usual bag model~\cite{bagmodel}. There the
bag describes the structure of the baryon containing {\it light\/}
quarks. The radius of the bag in this situation is determined by
the balance of the vacuum pressure and the pressure due to the
free motion of the light quarks inside, and in fact depends on the
quark wave function. In our case the inside of the bag contains
heavy quarks. Their kinetic energy is small, and we have treated
them here as static. The radius of the bag thus is determined
purely by the dynamics of the scalar glueball field and is not
sensitive to the state of the heavy quarks. This is true for low
lying excitations for which the radius of the quark state is
smaller than the inverse glueball mass. When these two scales are
comparable presumably the quark pressure will also be important
and will play a role in the energy balance. Thus in this
intermediate regime we expect the $V$-vortex to be similar to the
bag in the usual bag model. For states of even larger size the
potential between quarks is linear with the fundamental string
tension. The bag picture should therefore go smoothly into the
string picture.

Three space-time dimensions are unique in many ways. In particular they allow
massive gauge theories, where the mass of the gauge fields does not arise due
to the Higgs mechanism. We mean here the Chern-Simons gauge theories.
The discussion of the magnetic symmetry in 2+1 dimensions would not be
complete if we did not discuss this type of models. The next section is devoted to
such a discussion.

\section{Magnetic symmetry and the Chern-Simons term.}

The question of vortices in the Chern-Simons theories is interesting for the
following reason.
The Yang-Mills-Chern-Simons (YMCS) theories are completely massive, and so the Wilson loop
is expected to have a perimeter law behavior \cite{nair}. By the argument we have established
earlier, it should
then follow that the magnetic symmetry is unbroken and the spectrum should contain
vortex states.
This is an interesting proposition worth investigating.
This conclusion would not hold if the magnetic symmetry had no local order
parameter.
If that were the case the symmetry can be unbroken,
$W$ may have perimeter law and there could still be
no magnetic vortex states.
This is the main question we address in this section.
We want to examine more
carefully the question of locality of the vortex operator.\cite{Dunne}

We start our discussion by considering a simpler theory - compact
electrodynamics with CS term. Conceptual questions here are similar but the
technical side is much simpler. We find
that generically the theory does indeed contain local vortex operators
and a global discreet magnetic symmetry which is unbroken.
Nevertheless in the continuum limit there are no magnetic vortices in
the spectrum.
The reason is that the energy of such a vortex is logarithmically UV divergent.
We find however that with a particular scaling of the CS coefficient (logarithmically
vanishing when UV cutoff is removed) the energy of the vortex becomes finite. This
suggests that the theory may indeed have a phase with finite energy vortex states
and vanishing photon mass.

\subsection{Compact QED with the Chern Simons term.}
The Lagrangian of Abelian Chern Simons theory in the formal continuum limit is
\be
 \L =  -{1\over 4 g^2 } F_{\mu\nu} F^{\mu\nu}
 +{\kappa\over 2} \ep^{\mu\nu\rho} A_\mu \partial_\nu A_\rho
\label{Lagrangian1} \ee The gauge coupling $g^2$ has dimension of
mass and $\kappa$ is dimensionless. Equations of motion read as
$\partial_{\nu}F^{\mu\nu}=  \kappa g^2 \ep^{\mu\nu\rho}
\partial_\nu A_\rho$. The mass of the gauge particle is $M = \kappa
g^2$. The canonical structure of the theory is simplest in the
Hamiltonian gauge, $A_0= 0$.

The Hamiltonian is
\be
H= {1\over 2g^2}( E_i^2 +B^2)
\label{hamiltonian}
\ee
with canonical momenta related to the time derivatives of the fields by
$\Pi^i= -{1\over g^2} {\dot{A}}_i + {\kappa\over 2}\epsilon^{ij}A_j$.

The gauge fields and canonical momenta form the canonical algebra, and
the algebra involving the electric fields is
\be
[E_i(\vec{x}),E_j(\vec{y})] = -i \kappa g^4\epsilon_{ij}
\delta^2(\vec{x}- \vec{y}), \hskip 1 cm
[A_i(\vec{x}),E_j(\vec{y})] = -ig^2 \delta_{ij} \delta^2(\vec{x}- \vec{y})
\ee
The Gauss law,
\be
 \partial_iE^i -\kappa g^2 \epsilon^{ij}\partial_i A_j= 0
\ee
 on a spatial plane
generates time independent local gauge transformations. The
elements of the local gauge group take the form \be U(\lambda) =
\mbox{exp} \big\{{1\over g^2} i\int d^2x\,\lambda(x) \big(
\partial_iE^i - g^2\kappa \epsilon^{ij}\partial_i A_j \big)\big \}
\label{gausslaw} \ee such that $UAU^{-1} = A + d\lambda$. In the
non-compact theory $\lambda \in R$ should be a single-valued
function such that the eigenvalue of the operator (\ref{gausslaw})
on physical states is unity. Singular $\lambda$'s correspond to
transformations which are in general nontrivial on the physical
states.

Our interest however is in the compact theory. This means that
magnetic vortices of flux $2\pi$ must be physically unobservable.
As discussed in \cite{Kovner} this amounts to
further restricting the physical Hilbert space to states
which are trivial under the action of the vortex operator.

In other words
certain large gauge transformations must act on the physical states
trivially in the compact theory as opposed to the non-compact one.  The
compact gauge group therefore includes these
singular gauge transformations in addition to the regular ones,
which form the gauge group in the noncompact theory.
Consider
a multi-valued angle function
$\theta(x,x_0)$ which is singular at one point and
has a discontinuity along a straight
curve $C(x_0)$ that starts at the point $x_0$
and goes to infinity. The operator of the gauge transformation with this
singular gauge function creates a magnetic vortex.
Its explicit form (after partial integration and dropping a boundary term
owing to the
fact that all gauge invariant fields decay at infinity) is
\be
\tilde V(x_0) =
\mbox{exp} \{ -{1\over g^2} i\int d^2x\,{\tilde{\partial_i}}
\theta(x - x_0) \Bigg( E^i -g^2\kappa \epsilon^{ij} A_j \Bigg)\}
\label{vortex}
\ee
We have defined
\be
{\tilde{\partial_i}}\theta(x - x_0) = \partial_i\theta(x - x_0) -
2\pi\epsilon_{ij} c(x)_j \delta (x- C(x_0) ) =
 {\epsilon_{ij}(x- x_0)_j\over (x-x_0)^2}
\ee
$c(x)_j$ is a unit vector tangent to the curve $C(x,x_0)$.

We may want to include $\tilde V$ into the compact gauge group.
However to be part of the gauge group, it must commute (at least
weakly) with other elements of the group. One can check explicitly
that  $\tilde V(x,x_0)$ does not commute with the elements of the
noncompact group. To rectify this situation we define, following
\cite{Kovner} a slightly modified operator \be V(C, x_0)=
\mbox{exp}{2\pi i\over g^2}\epsilon_{ij}\int d^2 x c(x)_j \delta
(x- C(x,x_0))E_i(x) \ee This operator is  merely a ``collection''
of the electric fields which are perpendicular to the curve
$C(x_0)$. More explicitly one can write it in the following form
\be V(C, x_0)= \mbox{exp}{2\pi i\over g^2}\epsilon_{ij}\int_C
dl_i\,E_j(x) \label{vortex3} \ee Gauge invariance of this
operator, $[V, U]=0$, follows immediately. We also need to check
the commutativity of $V(x)$ with $V(y)$. A straightforward
calculation gives , \be V(C_0)V(C_1)= V(C_1) V(C_0){\mbox{exp}}
\Bigg \{i 8 \pi^2 \kappa L(C_0,C_1) \Bigg \} \ee Where
$L(C_1,C_2)= \pm 1$ if the curves cross each other and $ L= 0$ if
they don't. In order that $V$ be a Lorentz scalar the commutator
should not depend on the curves $C_1$ and $C_2$ To guarantee this
we need to set \be 4\pi\kappa = k \in Z \ee We find therefore that
the requirement of compactness quantizes the coefficient of the
Chern Simons term very much like in the non-Abelian theory .

The commutator \be [B(x), V^m(x_0)]= 2\pi m \delta^2(x-
x_0)V^m(x_0), \hskip 1 cm m \in Z. \ee indicates that
(\ref{vortex3}) creates magnetic vortices of integer strength.
Being gauge invariant this operator also creates an electric
charge \be Q= {1\over g^2} \int_\Sigma d^2 x \partial_i E^i =
{k\over 4\pi} \int_\Sigma d^2 x B = {mk\over 2} \ee Since $V$ has
to be included in the gauge group, the magnetic flux and the
electric charge created by it must be unobservable. Therefore the
Hamiltonian of the theory must commute with $V$. The noncompact
Hamiltonian eq.(\ref{hamiltonian}) does not quite do the job. It
should be modified but in such a way that in the continuum limit
the same form is recovered for smooth fields. The modified
Hamiltonian that satisfies these conditions has been suggested in
\cite{Kovner}. Since the UV structure is important for our
considerations, it is most usefully presented in the lattice
notations \be H_B= {1\over a^4 g^2 n^2}\sum_x\big( 1-
\mbox{Re}\,e^{i n a^2 B(x) } \big), \hskip 1 cm H_E= {m^2 g^2\over
4\pi^2 a^2 }\sum_x \big( 1- \mbox{Re}\,e^{i2\pi {a \over m g^2
}\epsilon_{ij}\hat{n}_j E_i(x)} \big), \label{latticehamiltonian}
\ee $a$ is the lattice spacing and $ m,n \in Z$ and $\hat{n}_j$ is
the unit vector parallel to the link. The normalization of the
electric and magnetic terms is such that in the naive continuum
limit $a\rightarrow 0$ they reduce to $B^2$ and $E^2$
respectively.

Using the Gauss' law one can see
that if $2n = k$ , the magnetic part
$H_B$ becomes a combination of the vortex operators $V$.
Therefore without loss of generality we assume $2n < k $.
For $m=1$ the electric part of the Hamiltonian is also a sum of a fundamental
vortex and anti-vortex, and we take $m > 1$.

Now that we have the formulation of the compact CS QED we can ask
about the locality properties of vortex operators. The operator
$V$ we have considered so far is of no interest of itself, since
is is trivial on all physical states. We thus have to look at the
operators which create magnetic flux smaller than $2\pi$ \be
V_p(C, x_0)= \mbox{exp}{2 p \pi i\over g^2}\epsilon_{ij}\int_C
dl_i\,E_j(x) \label{nonintegervortex},\hskip 1 cm p \in Q \ee Here
$p$ is a rational number $p \in (0,1)$. The question we are asking
is, are there such values of $p$ for which $V_p$ is a gauge
invariant local operator. The gauge invariance with respect to the
noncompact gauge group is straightforward, since $V_p$ only
depends on the electric field, and the electric field itself is
gauge invariant. However $V_p$ should also commute with the
"fundamental" vortex $V$, since $V$ is part of the gauge group.
Therefore we have \be [V,V_p]=0 \hskip 1 cm \Longrightarrow k p =
l \in  Z \hskip 1 cm \mbox{and}\hskip 0.3 cm l < k.
\label{condition1} \ee The condition of locality requires that
$V_p(x)$  commute with each other at different points $x$ and $y$.
This commutator also should be independent of the contour $C$ in
the definition eq.(\ref{nonintegervortex}). \be [V_p(x),V_p(y)]=0
\hskip 1 cm \Longrightarrow k p^2 = r \in  Z \hskip 1 cm
\label{condition2} \ee Both equations (\ref{condition1}) and
(\ref{condition2}) have to be satisfied for the existence of
non-trivial local vortices. Whether it is possible or not to
satisfy these equations clearly depends on the CS coefficient $k$.
For example there are no solutions for $k= 2$ and $k= 3$ theories.
For $k=4$ we can choose $l= 2$ and this gives a vortex of
vorticity $p = 1/2$. In general one can solve the constraints in
the following way. Writing CS coefficient in terms of its prime
factors, $k= q_1\,q_2\,q_3... \,\,q_m$, where all $q_i$ are not
necessarily different, one has the following two conditions to
satisfy \be q_1\,q_2\,q_3...\,q_m\, p = l, \hskip 1 cm
q_1\,q_2\,q_3...\,q_m\, p^2 = r. \ee The first condition is solved
if $p$ divides $k$, \be p =  {1\over q_1\,q_2\, q_3...\,q_i},
\hskip 0.5 cm \mbox{where} \hskip 0.5 cm i< m \ee Using this in
the second condition one can see that the most general form of $k$
which allows vortices is \be \mbox{If}\hskip 0.3 cm  k =  t^2 z
\hskip 0.5 cm \Longrightarrow  p = {1\over t},\hskip 0.5 cm t \geq
2 \hskip 0.3 cm \mbox{and} \hskip 0.3 cm t,z \in   Z \label{tz}
\ee For example if $k$ is a prime number there are no solutions.
Generically it is easier to find a solution at large values of
$k$.

The above relations also
show that should a solution exist, there is always a vortex of minimal vorticity.
All the other local vortices are simple powers of this minimal vortex.
For example for $k = 36$,
the above conditions give three solutions (and their integer multiples), $ p=(1/2, 1/3,1/6 )$.
Obviously ``p= 1/6'' is the minimal vortex.
The minimal value of $p={1\over w}$ determines the global magnetic symmetry group
of the theory as $Z_{w}$.

One last requirement that $V_p$ should satisfy, is locality with
respect to the energy density eq.(\ref{latticehamiltonian}). In
obvious notation \be [h_E(x), V_p(y)]=0, \ \ \ \ x\ne y
\Longrightarrow {kp\over m} \in   Z \ee This can always be
satisfied by choosing $m = k$. To satisfy the other
 condition
\be [h_B(x), V_p(y)]=0  \Longrightarrow {n p\over 2} \in Z \ee we
can take $n=2$. Certainly one can define other Hamiltonians which
will be compatible with the above conditions. We see therefore
that for those values of $k$ for which vortex operators are local
with respect to each other we can always choose the Hamiltonian
such that they are also local relative to the Hamiltonian density.

Thus we conclude that for many values of $k$ local physical vortex operators
exist.
They are order parameters for a global $Z_w$ magnetic symmetry. The value of $w$
is determined by $k$ through the solution of the equations for minimal $p$.
Thus the argument described above
 applies and, at least in the lattice theory
there are vortex states. Calculating their energy in the lattice theory
is not a simple matter. However
the interesting question is whether these states survive in the
continuum limit. That is to say, whether their energy
stays finite as the lattice spacing approaches zero.

In the continuum limit for smooth configurations of the fields
the theory is described by the Lagrangian eq.(\ref{Lagrangian1}).
However while solving continuum equations we may sometimes encounter
field configurations with fast variations. For these configurations it is important
to take into account the compactness of the theory.
In particular consider the electric field created by the
"minimal" vortex operator
$V_{1/w}$.
\be
[V(x),E_i(y)]=E_i(y)+e_i(x,y),\ \ \ \ e_i(x,y)=
{1\over w} g^4\kappa\hat n(y)_i \delta (y- C(x,y))
\label{field}
\ee
where $\hat n(y)$ is the vector tangential the curve $C$ at the point $y$.
Since the operator $V(x)$ is local,
its only observable action in the compact theory is at the point $x$.
However if we just calculate the energy using the naive Hamiltonian
eq.(\ref{hamiltonian})
we find infrared divergence proportional to the length of the curve $C$.
Clearly if faced with this type of configurations in continuum calculations
we should subtract this infrared divergence by hand.
Rather than do this we find it convenient to think about it in the
following way. Let us split the general electric field configuration
into a smooth piece and a piece that contains arbitrary number
of strings of the type of eq.(\ref{field})
\be
E^i=E^i_{smooth}+e^i
\ee
and subtract the contribution of $e_i$ in the Hamiltonian.
The only remnant of $e_i$ then is in the Gauss' law, since $e_i$ of
eq.(\ref{field}) corresponds to a pointlike charge $ {k\over 2w}$ at the
point $x$. Thus the smooth field
$E^i_{smooth}$ satisfies not the naive Gauss' law, but rather a modified one
\be
\partial_iE^i_{smooth}|_{{\rm mod}{k\over 2w}\delta^2(x)}
-\kappa g^2 \epsilon^{ij}\partial_i A_j= 0
\label{gaussviolation}
\ee
In other words we can work entirely in terms of $E_{smooth}$ if we remember that
we may allow Gauss' law to be violated by the presence of
pointlike charges of charge $\kappa g^2/w$. The appearance of $w$ in this way is
the only remnant of the compactness of the theory. In the following we will
work in terms of the smooth fields but will drop the subscript $smooth$ for
brevity.

With this caveat in mind,
to determine the energy of the magnetic vortex in the continuum limit
we now should solve the continuum equations of motion.
For a minimum vorticity solution $1/w$
following \cite{Nielsen, Khare} one can take the time independent symmetric
ansatz,
\be
A_i(r)=  \epsilon_{ij}{x_j\over r^2}[ g(r) - {1\over w}] \hskip 1 cm
A_0(r)=  h(r)
\ee
The equations of motion read
\be
g''(r) -{1\over r} g'(r) -rM h'(r) =0, \\
h''(r) +{1\over r} h'(r) -{M\over r} g'(r) = 0,
\label{gausslaw2}
\ee
where $ M= g^2\kappa$. We are looking for the solutions with
vorticity $1/w$. The magnetic field is $B = - {1\over r} g'(r)$ so we impose
$g(0)= 1/w$ and  $g(\infty)= 0$ and we also demand that the fields decay
exponentially at infinity. The solution under these conditions
can be found as
\begin{eqnarray}
&&g(r)= {Mr\over w}K_1(Mr), \hskip 1 cm B(r)= {M^2\over w} K_0(Mr) \\
&& h(r)= -{M\over w} K_0(Mr)\hskip 1cm
E^i= -{x^i  M^2\over wr}\,K_1(Mr),
\label{solution}
\end{eqnarray}
where $K_0, K_1$ are the Bessel functions.The energy of this vortex follows as
\be
{\cal{E}}= {\pi\over g^2} \int_0^\infty r\,dr \{ ({dh\over dr})^2 +
{1\over r^2} ({dg\over dr})^2 \}= {\pi\kappa\over w^2} M \big( -\gamma_E +
\mbox{ln}2 - \mbox{ln}{M\over \Lambda} \big)
\label{energy}
\ee
where $\Lambda$ is the ultraviolet cutoff scale.

This result warrants several comments. First, we see that the energy
of the vortex is
IR finite. This is closely related to the fact that the locality of the
operator $V_{1/w}$ allowed us to "violate" the Gauss' law.
Looking at the electric and magnetic fields in eqn (\ref{solution})
we indeed see that the naive Gauss' law is violated precisely by the amount
allowed by eq.(\ref{gaussviolation}).

Second, the energy is UV divergent. Thus the magnetic vortices do not
survive in the
continuum limit as finite energy excitations. This of course does not contradict
our original argument.
The $Z_{w}$ symmetry is unbroken in the vacuum and excitations carrying the
quantum numbers of this symmetry are very heavy.
The symmetry therefore seems
completely irrelevant for the low energy dynamics.
The curious thing though is that although the energy of the vortex is large,
it is still
much lower than the natural ultraviolet scale $\Lambda$. The vortices therefore
are not genuine ultraviolet objects in the lattice theory, but rather occupy
an intermediate scale between the UV scale $\Lambda$ and the IR scale M.
In fact bringing $M$ down to zero makes vortices light.
At $M=\mu\ln^{-1}{\Lambda\over \mu}$
the energy of the vortex is finite\footnote{The reader may wonder why we
are not bothered
by the factor $\kappa$ in eq.(\ref{energy}). After all we saw
that in the continuum limit naturally $\kappa\rightarrow\infty$. The point is that
the flux of the minimal vortex generically scales as $w^2\propto k $. For
example if $k=x^2$ with some integer $x$, then clearly the minimal
solution of the eq.(\ref{tz}) corresponds to $w=x$. Thus $k /w^2$ is
finite in the continuum limit.}.
This behavior in fact is very reminiscent of vortices in the
Higgs phase of the Abelian Higgs model.
The mass of such a vortex at weak coupling is
very large; $M\propto M_v^2/g^2\ln M_H/M_v$, where $M_v$ is the mass of the massive
photon and $M_H$ is the mass of the Higgs particle.
However as the photon mass decreases, that is as the
theory approaches the phase transition line, vortices become light.
On the phase transition itself they in fact become massless and condense
in the Coulomb phase.
It is not unlikely that similar phenomenon occurs in our model. As $\kappa$
decreases at
fixed $g^2$ the photon becomes lighter, and the mass of the vortex
also decreases. It could happen that at some value of $\kappa$ the vortices
actually become massless and drive a phase transition into a phase with broken
magnetic symmetry. A transition of
precisely such type was conjectured to happen in the lattice model
at $ k =8$ in \cite{Kogan}
and was seen in the variational calculation of \cite{Kovner}.
Of course, within the naive continuum limit we consider here
we are unable to see such behavior. However the fact that the
vortices
 become
light within the validity of the naive continuum limit is quite suggestive
in this respect.
Note that if the vortices indeed do condense, the
magnetic $Z_w$ symmetry is spontaneously broken. By virtue of the
argument given in Sec. 3 this means that the low $k$ phase
is confining.

\subsection{Non-Abelian theories.}

We now extend this analysis to non-Abelian $SU(N)$ theories, \be
\L = {1\over 2g^2}\mbox{tr}F_{\mu \nu}F^{\mu \nu}
-\kappa\epsilon^{\mu \nu \lambda}\mbox{tr}\left(A_\mu \partial_\nu
A_\lambda +{2\over 3} A_\mu A_\nu A_\lambda \right) \ee CS
coefficient has the well known quantization, $4\pi\kappa= k$
\cite{Deser}. The classical equations of motion follow as \be
D_\mu F^{\mu\nu} -{1\over 2}g^2 \kappa
\epsilon^{\nu\lambda\mu}F_{\lambda \mu} = 0 \ee The canonical
structure of this theory is similar to the Abelian case. In the
$A_0^a= 0$ gauge \cite{dunne} \be \Pi^a_i= -{1\over g^2} E_i^a +
{\kappa \over 2}\epsilon^{ij}A_j^a, \hskip 1 cm \mbox{where}
\hskip 1 cm    E_i^a = \dot{A}_i^a. \ee The Hamiltonian is \be H=
{1\over 2g^2}\{ (E^a_i)^2 +(B^a)^2 \}, \hskip 1 cm \mbox{where}
\hskip 1 cm B^a =  {1\over 2}\epsilon_{ij}\, F^a_{ij}.
\label{hamiltonian2} \ee The canonical algebra is \be
[E_i^a(\vec{x}),E_j^b(\vec{y})] = -i\delta^{ab} \kappa
g^4\epsilon_{ij} \delta^2(\vec{x}- \vec{y}), \hskip 1 cm
[A_i^a(\vec{x}),E_j^b(\vec{y})] = -ig^2\delta^{ab} \delta_{ij}
\delta^2(\vec{x}- \vec{y}) \ee In terms of the momenta the Gauss
law is \be
 (D_i\Pi_i)^a = - {\kappa\over 2} \epsilon^{ij}\partial_i A_j^a
\ee

In the non Abelian YMCS theory, the large Wilson loop still
commutes with the Hamiltonian. This is obvious in the Hamiltonian formalism,
since the commutation relation between the vector potential and the
chromoelectric field
is unaffected by the presence of the CS term. The form of the Hamiltonian
in terms of $A_i$ and $E_i$ is also the same as without the CS term.
Since $W$ is a function of $A_i$ only, its commutator
with the Hamiltonian is exactly the same as in the theory without CS.
Therefore the fundamental Wilson loop still generates a symmetry.

Again our question is whether the theory admits local vortex operators.
In the non-Abelian theory our choices are more limited than in compact
QED. In the SU(N) Yang Mills theory, the only candidates for local operators
are those that create quantized flux \cite{thooft,kovner1}. The vortex operator
in YM theory is \cite{kovner1,kovner3}:
\begin{equation}
V(x)=\exp\{{ 4\pi i\over g^2N} \int_C dy^i
\epsilon_{ij}{\rm Tr}(YE_i(y))
\label{v21}
\end{equation}
where the hypercharge generator $Y$ is
\begin{equation}
Y={\rm diag} \left(1,1,...,-(N-1)\right)
\end{equation}
We have proved in the earlier sections that in SU(N) YM theory, this operator despite its nonlocal and
gauge non-invariant appearance is in fact a local, gauge invariant, Lorentz scalar field
\cite{kovner1,kovner3}. The way it was constructed there was to require that it satisfies
the 'tHooft algebra \cite{thooft} with the fundamental Wilson loop
\be
V^\dagger(x)W(C)V(x)=\exp\{{ 2\pi i\over N}n(x,C)\}W(C)
\label{th2}
\ee
with $n(x,C)$ being the linking number on the plane between the point $x$
and the closed curve $C$.

One can see that the operator in (\ref{v21}) is also the
appropriate vortex operator when a Chern-Simons term is included
for the gauge field. The commutation relation (\ref{th2}) is still
satisfied by the expression (\ref{v21}). However an additional
requirement is that $V$ be gauge invariant, in the sense that the
matrix elements of $V$  between physical states (those that
satisfy Gauss' law) and non-physical states (non-singlet under
gauge transformations) vanish. This means that when we calculate
matrix elements of any number of operators $V$ We thus have to
check that it transforms a physical state into another physical
state. The difference with the pure YM theory is that the wave
function of a gauge invariant state does not depend only on the
Wilson loops. The physical wave function should satisfy the
following equation \be i(D_i{\delta\over\delta  A_i})^a \Psi[A]=
{\kappa\over 2}\epsilon_{ij}\partial_i A_j^a \Psi[A] \ee The
general form of $\Psi$ has been determined in \cite{nair} in terms
of certain nonlinear variables. For our purposes we find it more
convenient to work directly in terms of the vector potentials
$A_i$. Let us take $\Psi$ in the form \be \Psi = \exp\{- i S\} \ee
The eikonal $S$ satisfies a linear inhomogeneous equation \be
D_i^{ab}{\delta\over\delta  A_i^b}S[A]= {\kappa\over
2}\epsilon_{ij}\partial_i A_j^a \label{ds} \ee The solution of the
homogeneous equation is indeed any functional that depends on
Wilson loops $S_0[W]$. We can find a particular solution of the
inhomogeneous equation using the following argument. $S[A]$ must
be a functional whose change under a standard gauge transformation
of vector potentials $\delta A_i=D_i\lambda$ is proportional to
${\kappa\over 2} \int d^2x \epsilon_{ij}\partial_i A_j^a\lambda^a
$. Such a functional can be represented as a Chern Simons action
on a space with a boundary. Let us introduce an additional
coordinate $\tau\in [-\infty,1]$ and functions of three
coordinates $A_i(x,\tau)$ so that at the boundary $\tau=1$, the
value of these functions is equal to the value of the vector
potentials in our theory $A_i(x,\tau=1)=A_i(x)$. Let us write the
Chern Simons term (in the Weyl gauge) on this manifold \be
S_{CS}=\int_{-\infty}^1d\tau\int d^2x
\epsilon_{ij}A_i^a(x,\tau)\dot A_j^a(x,\tau) \label{cs} \ee Under
the $\tau$ independent gauge transformation this action changes by
a boundary term \be \delta S_{CS}=-\int
d^2x\,\epsilon_{ij}\lambda^a(x)\partial_iA^a_j(x,\tau=1)= -\int
d^2x\,\epsilon_{ij}\lambda^a(x)\partial_iA^a_j(x) \ee which is
precisely of the form required to satisfy eq.(\ref{ds}). A
particular solution of eq.(\ref{ds}) is therefore \be S_p=
-{\kappa\over 2} S_{CS}= -{\kappa\over 2}\int_{-\infty}^1
d\tau\int d^2x \epsilon_{ij}A_i^a(x,\tau)\dot{A}_j^a(x,\tau) \ee
The introduction of the extra coordinate $\tau$ and the expression
eq.(\ref{cs}) is not at all unnatural. One should view this extra
coordinate as parameterizing a curve in the field space. With this
interpretation we have \be d\tau\dot A_j(x,\tau)=\delta A_i \ee
and \be \int_{-\infty}^1d\tau\int d^2x
\epsilon_{ij}A^i(x,\tau)\dot A_j(x,\tau)= \int_{\cal C} \delta
A_i\epsilon_{ij}A_j \ee where the line integral is taken over the
trajectory ${\cal C}$ in the field space which ends at the point
$\{A_i(x)\}$.

We have thus determined the general form of the wave function
of a physical state in the YMCS theory to be
\be
\Psi[A]=\exp\{ i{\kappa\over 2}\int_{-\infty}^1d\tau\int d^2x
\epsilon_{ij}A^i(x,\tau)\dot A_j(x,\tau)\}\Psi_0[W]
\ee
Now it is straightforward to
see how the vortex operator acts on it.
Under the action of the vortex operator
\be
V(x)\,A_i^a(y,\tau)\,V^{\dagger}(x) =  A_i(x,\tau) +
{4\pi\over N}\mbox{Tr}YT^a  \epsilon_{ij}\int dz_j \delta^2 (z- C(x,y))
\ee
Remembering that $E_i^a= -g^2 \Pi_i^a + g^2 \kappa\epsilon_{ij}A_j^a$,
we see that
the change in the phase factor in the wave functional is exactly cancelled by
the $A$-dependent term in the vortex operator eq.(\ref{v2}).
\be
V(x)\,e^{-i S_p}\,V^{\dagger}(x)= e^{-i S_p}.
\ee
Thus
\be
V\Psi[A]
=\exp\{i{\kappa\over 2} \int_{-\infty}^1d\tau\int d^2x
\epsilon_{ij}A^i(x,\tau)\dot A_j(x,\tau)\}\Psi_0[V^\dagger WV]
\ee
Clearly a gauge transformed vortex operator $V_\Omega$ has exactly the
same action on the
wave functional $\Psi[A]$,
\be
V\Psi=V_\Omega\Psi
\ee
which establishes gauge invariance of $V$ in the same sense as in the
YM theory.

We now can check the locality of the operator $V$ by calculating
straightforwardly the relevant commutation relation.
A simple calculation gives  $[V(x), V(y)]= 0$.
Thus the operators are local with respect to each other. When
considering the locality with respect to the Hamiltonian density we
are faced with the same ambiguity as in the Abelian theory. The
electric part of naive
continuum Hamiltonian is not local relative to $V$, since $E_i$ is
shifted by the action of $V$ along the curve $C$. Just like in
the Abelian case one should consider a properly regularized version of
$H$ in order to be able to draw a definite conclusion. In the
non-Abelian case such a regularized Hamiltonian is not
available. However in the Abelian case we saw that there is quite a
lot of flexibility in defining such a regularized version. In
particular we saw that whenever the vortex operators were local with
respect to each other, we were always able to define the local Hamiltonian
density. We expect that this situation persists
in the non-Abelian theory too.

The situation in the continuum limit is again similar to the Abelian
case.
There are no finite energy solutions of the non-Abelian equations of
motion which have finite vorticity. The only way to find such
solutions would be again to relax the Gauss' law constraint by
allowing point like charges which correspond to the singular
chromoelectric
field created by $V$. However again those IR finite configurations
will have UV logarithmically divergent energy.
In fact taking Abelian ansatz the YM equations of motion reduce to
those we considered in the previous section and thus lead to the same
energy dependence on the UV cutoff.
Strictly speaking this conclusion is only valid for large enough value
of $k$, since for small $k$ quantum corrections to this classical
analysis may be large. Thus again it is possible that at small $k$ the
theory is in a different phase as suggested in \cite{cornwall}.

To summarize, we have found that compact CS QED does admit local vortex operators
for many values of the CS coefficient $k$. The energy of the vortex
excitations however generically is
logarithmically UV divergent in the continuum
limit. With a particular scaling of the CS coefficient these
vortices become light and might condense at small values of $k$.
Our results for non-Abelian CSYM theory are similar. Local vortex
operators exist, but the particles that carry vorticity are heavy in the continuum limit.

These results are broadly compatible with suggestions made in the
literature that at low values of the Chern Simons coefficient the YMCS
theory might undergo a phase transition. If this happens it is very
likely that this other
phase has a broken magnetic symmetry and is therefore confining.
This is a very interesting possibility which seems worthwhile
exploring.

\section{Summary}

In this review we have discussed various aspects of the magnetic
symmetry in 2+1 dimensional gauge theories. This symmetry is
sometimes called topological and is looked at as a somewhat
esoteric concept. In fact, as we have discussed here, it is not
topological at all, in the sense that its charge density (or group
element per unit volume in the discrete case) is a gauge invariant
local operator. Moreover in gauge theories without matter fields
in the fundamental representation it has a local order parameter.
It thus behaves in all respects as a normal nontopological
symmetry. As such its mode of realization is rigidly linked to
confining properties of the theory. Whenever this symmetry is
spontaneously broken, the gauge theory is confining, while
unbroken realization of magnetic symmetry signifies existence of
massive gauge bosons.

We have constructed the low energy description of gauge theories based
on the effective Lagrangian for the order parameter of the magnetic symmetry. This
Lagrangian realises confinement in a very straightforward and simple way on the
classical level.

In theories with fundamental matter, the magnetic symmetry is
still present, but it does not have a local order parameter. As a
result it is implemented as a {\it local} rather than {\it global}
symmetry. We have shown how to implement it in a consistent way in
the low energy effective Lagrangian in the case of heavy
fundamental matter. Interestingly, the resulting effective
Lagrangian provides a bag like description of baryons, as quarks
surrounded by a "bag" of nonconfining vacuum.

In Chern-Simons theories we proved that the magnetic symmetry is also present and is
unbroken in the vacuum.
Magnetic vortices - the excitations carrying the magnetic charge are however
infinitely
heavy in the continuum limit, at least as long as the Chern-Simons
coefficient is large. For small coefficient there is a possibility that the
vortices become massless and condense, thus driving the theory into the confining
phase.

In short, magnetic symmetry is a "trademark" of 2+1 dimensional
gauge theories, and most qualitative features of their low energy
dynamics is determined by the realization of this symmetry. It is
hard to over emphasise the importance of symmetries, especially in
strongly interacting theories, where a symmetry may be our only
lead in understanding dynamics. Even more so in pure gauge
theories, where the magnetic symmetry is {\it the only} global
symmetry of which existence we know. Hopefully we will learn how
to utilize it to the fullest and will thus gain more information
about the strongly coupled regime of these theories.

\section*{Acknowledgments}
This work was supported by PPARC. I thank  G. Dunne,
C. Fosco, I. Kogan, C. Korthals-Altes, B. Tekin
 and especially B. Rosenstein for collaboration over
the years, which shaped my understanding of the issues discussed in this review.

\end{document}